\documentstyle[aas2pp4,psbox]{article}
\newcommand{\ulumi}{~erg~s$^{-1}$}
\newcommand{\uflux}{~erg~s$^{-1}$~cm$^{-2}$}

\slugcomment{Submitted to ApJ Suppl.}
\lefthead{J. Yokogawa et al.}
\righthead{Populations of the X-ray Sources in the SMC}

\begin{document}
\title{A Study of the Populations of X-ray Sources 
in the Small Magellanic Cloud with ASCA}

\author{Jun~Yokogawa\altaffilmark{1}, Kensuke~Imanishi, 
Masahiro~Tsujimoto, Mamiko~Nishiuchi\altaffilmark{1},
and Katsuji~Koyama\altaffilmark{2}}
\affil{Department of Physics, Graduate School of Science, Kyoto University, Sakyo-ku, Kyoto, 606-8502, Japan;
jun@cr.scphys.kyoto-u.ac.jp,
kensuke@cr.scphys.kyoto-u.ac.jp, tsujimot@cr.scphys.kyoto-u.ac.jp,
mamiko@cr.scphys.kyoto-u.ac.jp, koyama@cr.scphys.kyoto-u.ac.jp}

\author{Fumiaki~Nagase}
\affil{Institute of Space and Astronautical Science, Yoshinodai, Sagamihara, Kanagawa 229-8510, Japan; nagase@astro.isas.ac.jp}

\and

\author{Robin~H.~D.~Corbet}
\affil{Code 662, NASA/Goddard Space Flight Center, Greenbelt, MD 20771, U.S.A.; Universities Space Research Association;
corbet@bastet.gsfc.nasa.gov}

\altaffiltext{1}{Research Fellow of the Japan Society for the Promotion of Science (JSPS)}
\altaffiltext{2}{CREST, Japan Science and Technology Corporation (JST), 4-1-8 Honmachi, Kawaguchi, Saitama, 332-0012, Japan}

\begin{abstract}
The Advanced Satellite for Cosmology and Astrophysics (ASCA) has made
multiple observations of the Small Magellanic Cloud (SMC).
 X-ray mosaic images in the soft (0.7--2.0~keV) and hard (2.0--7.0~keV)
bands are separately constructed, and the latter provides the first
hard X-ray view of the SMC.
 We extract 39 sources from the two-band images with a criterion of $S/N>5$,
and conduct
timing and spectral analyses for all of these sources.
Coherent pulsations are detected from 12 X-ray sources;
five of which are new discoveries.
Most of the 12 X-ray pulsars
are found to exhibit long-term flux variabilities,
hence they are likely to be X-ray binary pulsars (XBPs).
On the other hand,
we classify
four supernova remnants (SNRs) as thermal SNRs,
because their spectra exhibit emission lines
from highly ionized atoms.
 We find that XBPs and thermal SNRs in the SMC
can be clearly separated by their hardness ratio
(the ratio of the count rate between the hard and soft bands).
Using this empirical grouping, we find many XBP candidates
in the SMC, although no pulsations have yet been
detected from these sources.
 Possible implications on the star-formation history and
evolution of the SMC are presented by a
comparison of the source populations in the SMC and our Galaxy.
\end{abstract}


\section{Introduction}
 Bright X-ray sources such as supernova remnants (SNRs) and binaries
with a neutron star (NS) or a black hole component are `relics' of
massive stars and hence carry key information on the
star formation history, dynamics, and structure of their host galaxy.
 However, our knowledge of the populations of bright X-ray sources in our Galaxy
is highly incomplete due to the Galactic absorption in
the low energy band,
large angular size that must be observed,
and limited information on the distance of the relevant
objects.

 The Small Magellanic Cloud (SMC), a satellite of
our Galaxy, is the next nearest neighbor after the Large Magellanic Cloud (LMC).
The proximity
(60~kpc is assumed in this paper; Mathewson 1985\markcite{Mathewson1985a}),
reasonable angular size
($\sim 3^\circ \times 3^\circ$),
and low interstellar absorption to the SMC are all favorable for an unbiased
survey of the X-ray source populations in the entire galaxy.
Surveys of soft X-ray sources (below $\sim 2$~keV)
have been carried out with the Einstein and ROSAT satellites.
The most complete source catalogs are
presented in Wang \& Wu (1992, hereafter WW92\markcite{Wang1992})
and Kahabka et al. (1999, hereafter K99\markcite{Kahabka1999}).
K99 classified the sources detected by the ROSAT PSPC
(Position Sensitive Proportional Counter),
based on two different hardness
ratios, together with spatial information.
Cowley et al. (1997)\markcite{Cowley1997} and 
Schmidtke et al. (1999)\markcite{Schmidtke1999}
made numerous observations of the SMC with the ROSAT HRI
(High Resolution Imager),
and determined accurate positions of X-ray sources.
Optical observations made at CTIO
(Cerro Tololo Interamerican Observatory) showed
that five of these X-ray sources have a massive star companion and
hence are high mass X-ray binaries (HMXBs).

 Bright X-ray sources can be classified by their spectral and temporal
characteristics. X-rays from young SNRs are either due to a shock
heated plasma (in shell-like SNRs) which preferentially emits soft
(below $\sim 2$~keV) line-dominated X-rays, or due to the release
of rotational energy from isolated NSs (X-ray pulsars associated with
Crab-like SNRs) with hard and featureless spectra.
Unlike SNRs, the other classes of X-ray sources emit time-variable
X-rays. NSs with a low mass companion star (low mass X-ray binaries:
LMXBs) have hard spectra and often exhibit X-ray bursts. Black hole
binaries (BHs) have bimodal behavior and show both
high-soft and low-hard spectral
states.
NSs with a high mass stellar companion
(HMXBs) have even harder X-ray spectra
and often exhibit long-term flux variations
of a factor $\gtrsim 10$--100.
Many HMXBs also show coherent pulsations (X-ray binary pulsars; XBPs).
 Consequently,
most bright X-ray sources emit
copious hard X-rays (above $\sim 2$ keV).
However, due mainly to a lack of imaging capability
in the hard X-ray band,
no systematic survey of the populations of hard X-ray sources
had previously been made.
The Advanced Satellite for Cosmology and Astrophysics
(ASCA; Tanaka, Inoue, \& Holt 1993\markcite{Tanaka1993})
is equipped with wide-band X-ray imaging instruments and
hence is optimized for an X-ray population study.

 In this paper, we report the results
from multiple observations on the SMC with ASCA.
The observation fields and the method of data reduction are
presented in \S\ref{sec:obs}.
In \S\ref{sec:result},
we extract X-ray sources detected above the 5-$\sigma$ level
from the observation fields,
perform temporal and spectral analyses on all the sources,
and categorize them into thermal SNRs and XBPs.
\S\ref{sec:comments} gives details of
individual X-ray pulsars and SNRs.
In \S\ref{sec:dis}, we propose a simple and reliable
source classification method.
We clearly distinguish between thermal SNRs and XBPs
with this method, then
discuss these X-ray source populations in the SMC.
Finally, we summarize this study in \S\ref{sec:conc}.

\section{Observations and Data Reduction \label{sec:obs}}
 ASCA
had observed the SMC 10 times by the end of 1998 and
we have used the data from eight of these observations which are
either in the public archive or came from our own proposed observations.
Observation dates, pointing directions and other relevant information
are listed in
Table \ref{tab:obs}.
Most of the optically bright portion and the Eastern Wing of
the SMC
has been covered.
In each observation, X-ray photons were collected with
four XRTs (X-Ray Telescopes) and detected separately with
the two GISs (Gas Imaging Spectrometers) and
the two SISs (Solid-state Imaging Spectrometers).
Details of these instruments are presented separately
in Serlemitsos et al. (1995)\markcite{Serlemitsos1995},
Ohashi et al. (1996)\markcite{Ohashi1996}, 
and Burke et al. (1994)\markcite{Burke1994}.

\placetable{tab:obs}

We first screened the X-ray data from both the GIS and SIS
with the normal criteria:
i.e. we rejected the data obtained in the SAA (South Atlantic Anomaly),
those with a cut-off rigidity lower than
4~GV,
and those with an elevation angle lower than $5^\circ$.
We then rejected the particle events in the GIS with a rise-time
discrimination technique, and the SIS data obtained when the elevation
angle from the bright earth was lower than $25^\circ$.
 The effects of RDD (Residual Dark Distribution) in the SIS data
were corrected if the observation was carried out later than 1996
with the method given in
Dotani et al. (1997)\markcite{Dotani1997}.
After the screening, the
total available exposure time of two GISs was $\sim570$~ks.

The GIS is more suitable than the SIS for X-ray population studies,
because of
its larger field of view, larger effective area at high energy,
and better time resolution.
This paper thus mainly utilizes the GIS data,
except for particular objects which require
high energy resolution and soft X-ray sensitivity.

 Photon events for individual X-ray sources were extracted from a circle of $3^\prime$ radius,
in which $\sim 90$\% of incident photons are contained
(Serlemitsos et al. 1995\markcite{Serlemitsos1995}),
or an ellipse for sources located significantly off-axis
because of the distortion of the point spread function of the ASCA XRTs.
Background data
were taken from a nearby off-source area for each source.
 In some source-crowded areas, we selected a smaller circle
with a radius $\sim 1\farcm5$--$2^\prime$ to minimize
cross talk from nearby sources.

\section{Analyses and Results \label{sec:result}}
\subsection{X-ray Images \label{subsec:image}}
\begin{figure}
\psbox[xsize=0.47\textwidth]{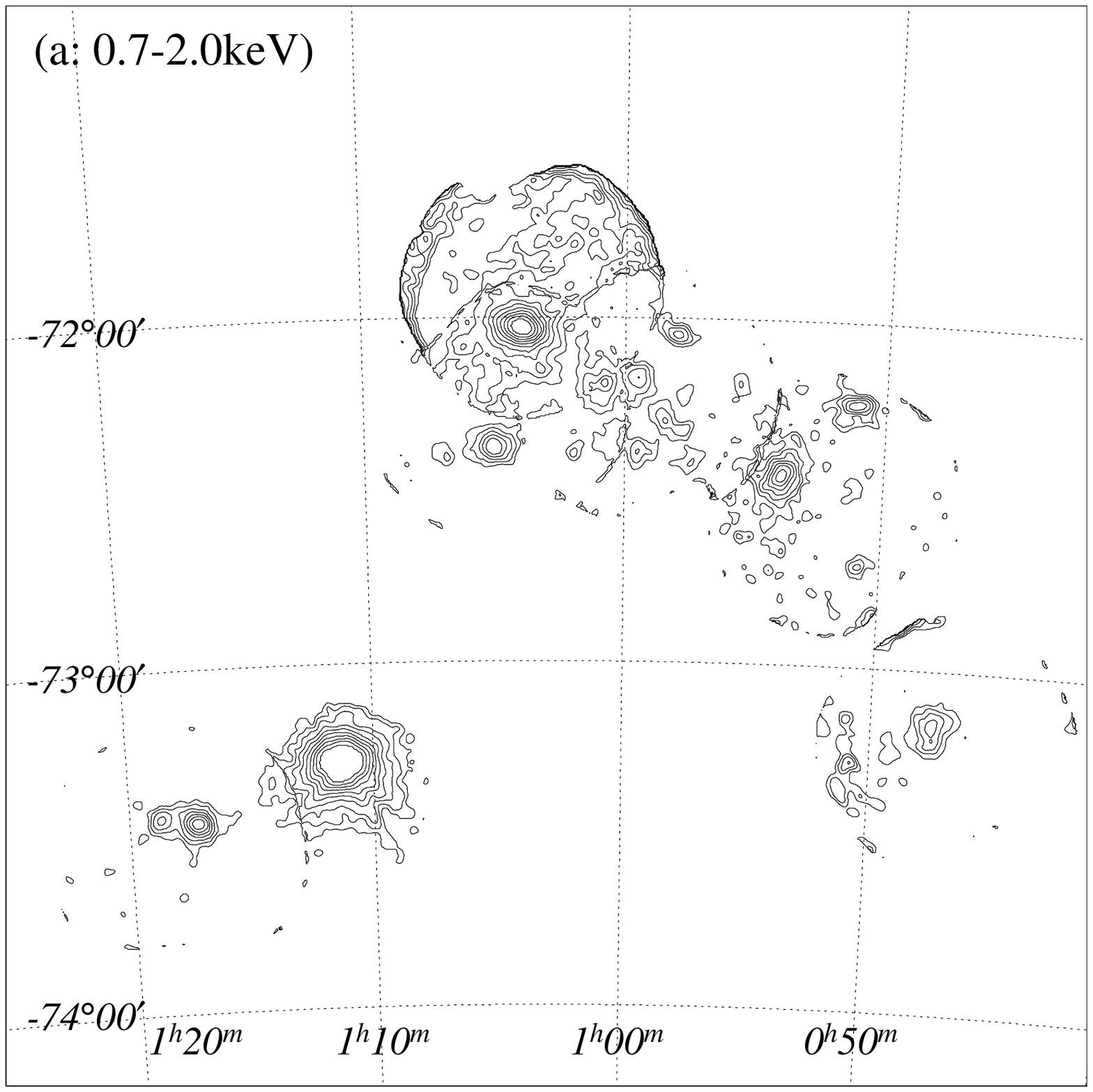}
\psbox[xsize=0.47\textwidth]{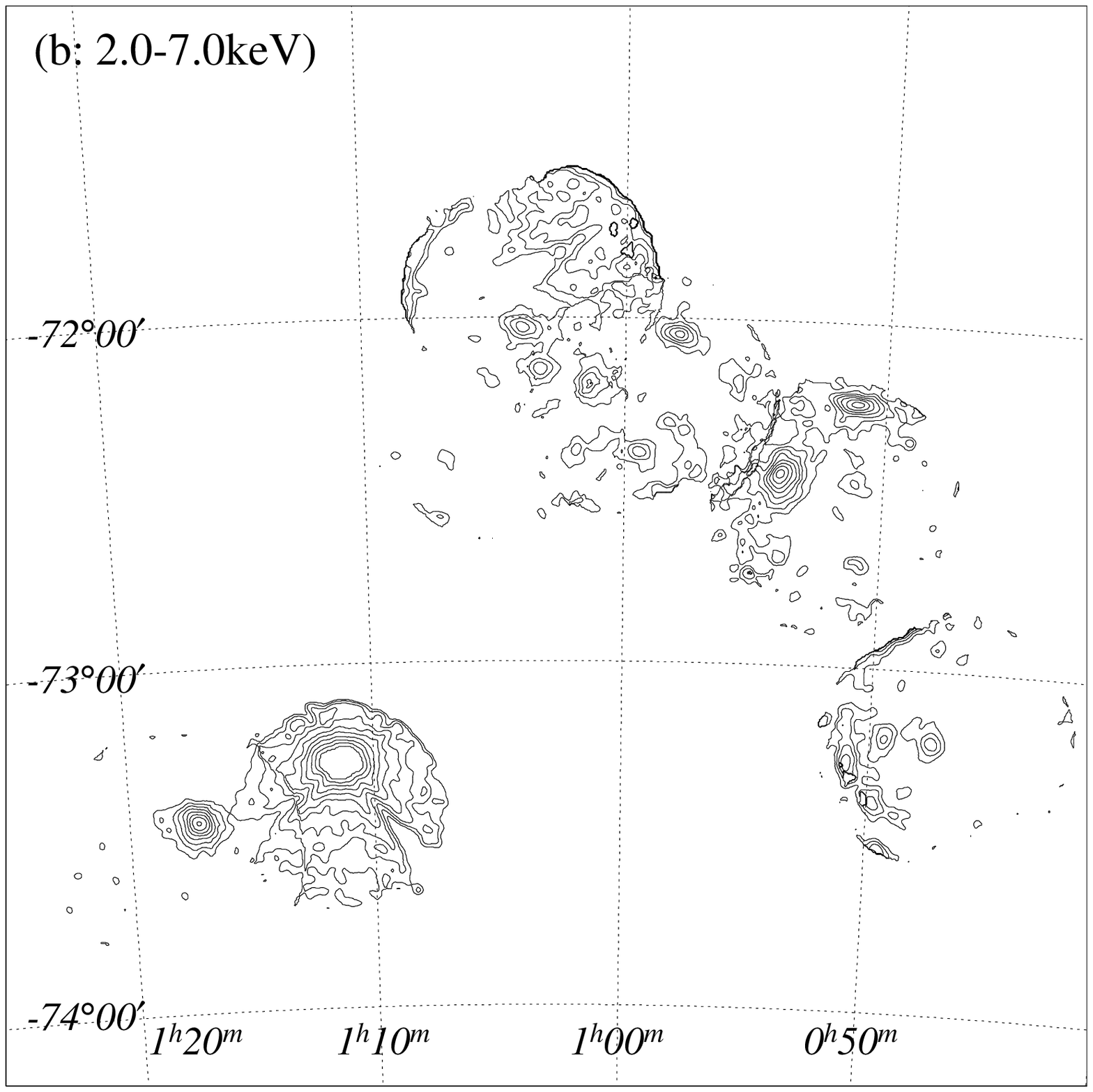}
\caption{SMC mosaic contour images obtained with ASCA GIS in
the soft (a: 0.7--2.0~keV)
and hard band (b: 2.0--7.0~keV),
overlaid with
the equatorial coordinates (J2000).
The effects of non X-ray background,
telescope vignetting and
difference of exposure time between observations
were corrected.
The complex structure for sources near the edge of the detector fields is due
to the point spread function of ASCA XRTs.
Contour levels are logarithmically
spaced and highly saturated at XTE J0111.2$-$7317.
\label{fig:smc}
}
\end{figure}

Figure \ref{fig:smc} shows the
mosaic X-ray images in the soft (0.7--2.0~keV)
and hard (2.0--7.0~keV) bands constructed from the multiple observations
and
the telescope vignetting,
non X-ray background and exposure differences have
been corrected.
 Of the two observations centered on SMC X-1,
only one (observation C),
when SMC X-1 was much dimmer,
was used.
The ASCA soft band image of the SMC is different from
those obtained with Einstein and ROSAT (WW92; K99),
indicating that many discrete sources in the SMC
are highly variable.
The hard band image is the first such image obtained
for the SMC
and it is found to have significant differences from the soft band image.

\placefigure{fig:smc}

\subsection{Source Catalog \label{subsec:cat}}

X-ray sources were identified from the soft/hard band images
from each observation,
with the criterion that the $S/N$ ratio should exceed 5-$\sigma$
in at least one of the two-band images.
Consequently 39 sources were found and are listed
in Table \ref{tab:smccat}.
The separation angle $\sim 2^\prime$ between sources No.\,4
and No.\,5
is comparable to the half power radius of the XRT,
hence we estimated the $S/N$ ratio in a circle
containing both of them and found it exceed 5-$\sigma$.
 Since No.\,4 appeared to be prominent in the soft band
while No.\,5 was strong in the hard band, we are confident
that there really exist two separate soft (No.\,4) and hard (No.\,5) sources.
The $S/N$ ratio of No.\,34 exceeds 5-$\sigma$ in the hard band,
while in the soft band it suffers severe contamination from No.\,30.

We derived hardness ratios (HRs) for all the sources
except for No.\,4, No.\,5, and No.\,34,
and cataloged them in Table \ref{tab:smccat}.
The HR is defined as ${\rm HR} = (H-S)/(H+S)$,
where $H$ and $S$ are the background subtracted GIS count rates
in the 2.0--7.0~keV and 0.7--2.0~keV bands respectively.

\placetable{tab:smccat}
\notetoeditor{Please place the first half of Table \ref{tab:smccat} 
on an even page and the second half on the next odd page, 
so that Table \ref{tab:smccat} covers the two facing pages. }

\subsection{Timing Analysis \label{sec:timing}}

\begin{figure}
\psbox[xsize=0.47\textwidth]{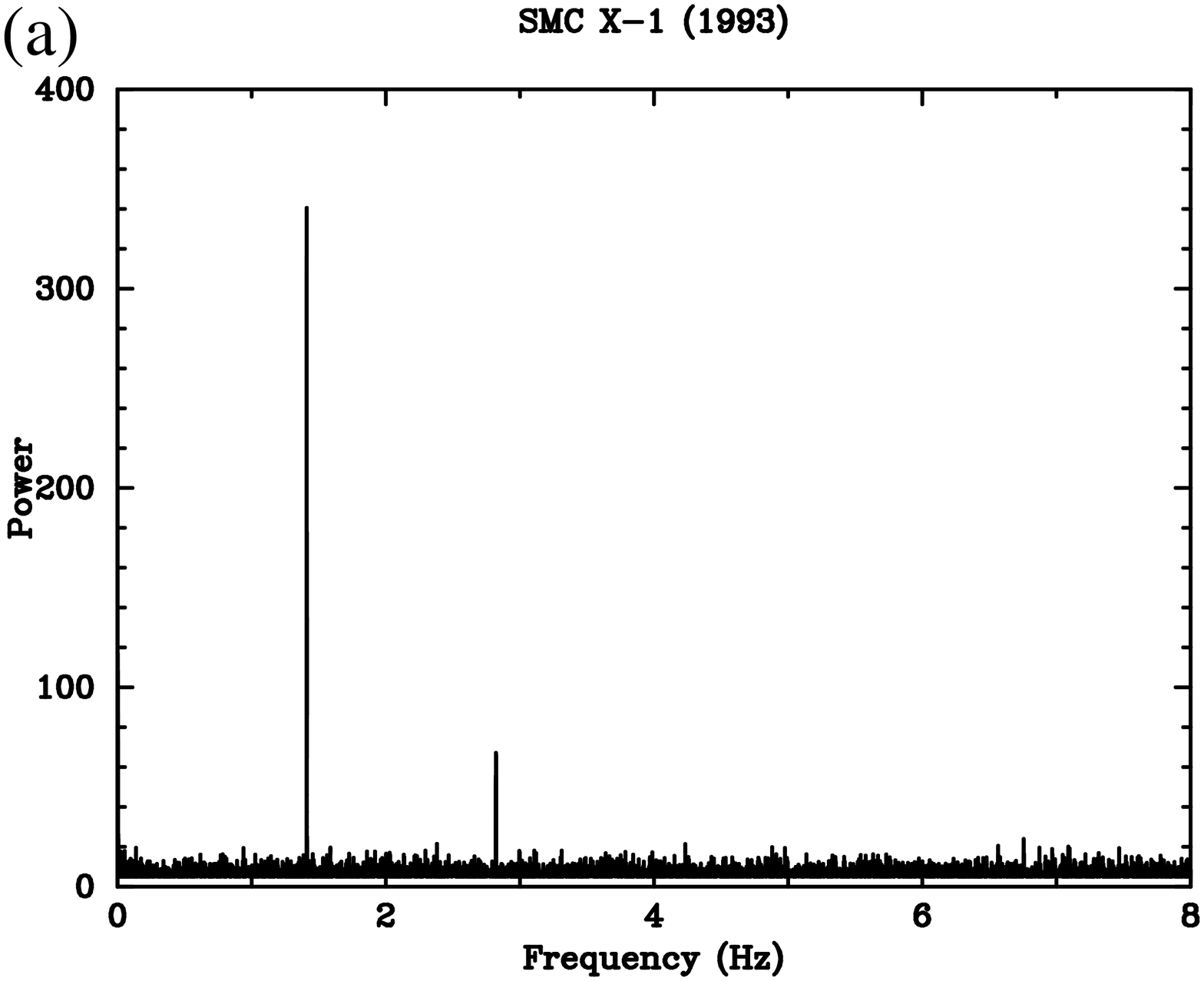}
\psbox[xsize=0.47\textwidth]{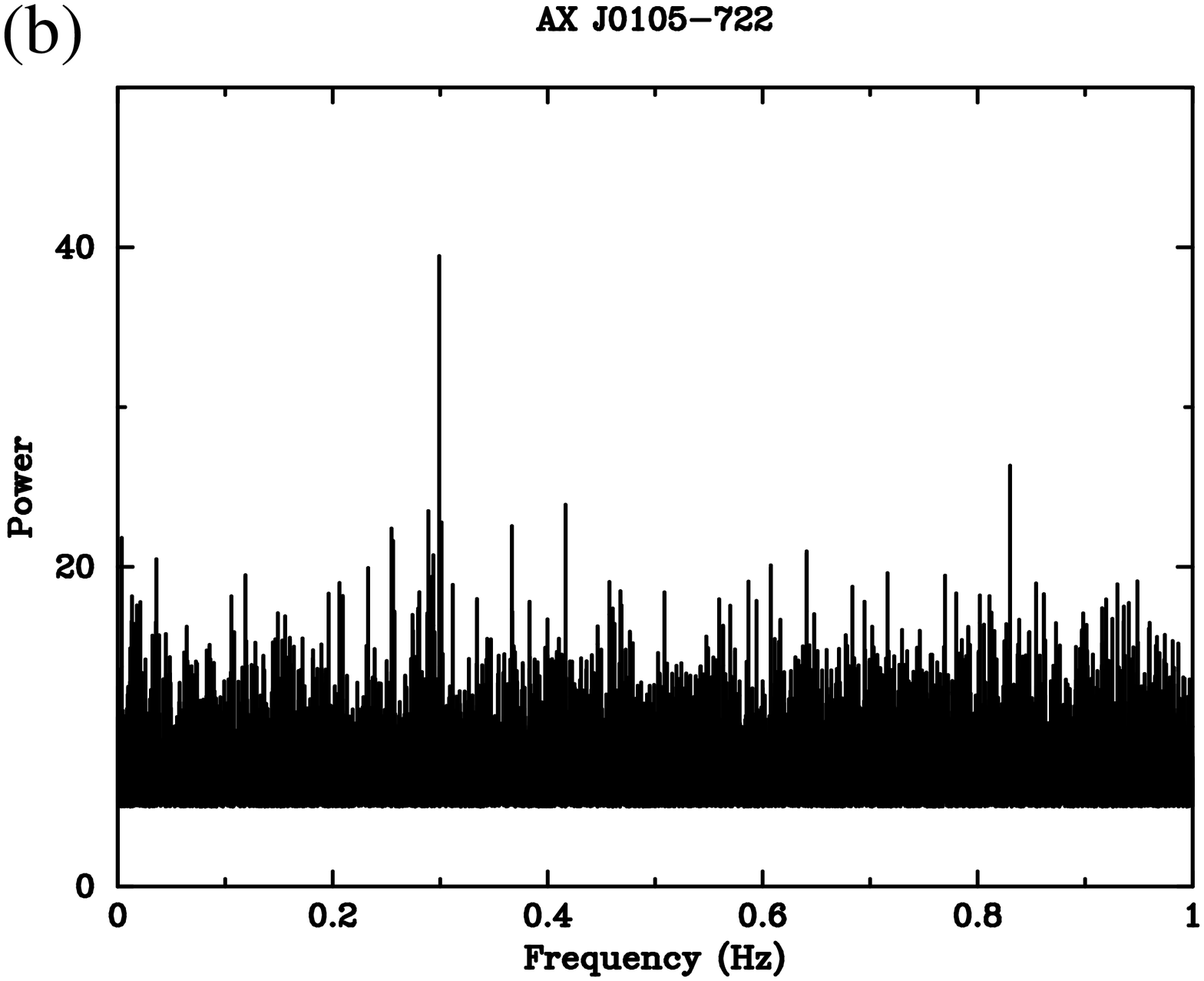}
\caption{Power spectra of SMC X-1 in observation A (a) and AX J0105$-$722 (b), 
where data points larger than 5 are plotted. 
Pulsations were detected unambiguously from most of pulsars like (a), 
while AX J0105$-$722 exhibited rather weak evidence as shown in (b). 
\label{fig:psd}
}
\end{figure}

 We performed an FFT (Fast Fourier Transform) analysis
on all the sources to search for coherent pulsations.
We used only high-bit rate data in order to utilize
the maximum time resolution (up to 62.5~ms)
for those sources with high count rates.
Otherwise
we used high-bit and medium-bit data simultaneously
to achieve better statistics
at the sacrifice of the time resolution reduced to 0.5~s. 
We detected coherent pulsations from 11 sources,
five of which are new discoveries from this study.
We show a power spectrum obtained from SMC X-1 (No.\,38; in 
observation A) in Figure \ref{fig:psd} (a) as an example of 
unambiguous detection. 
Only AX J0105$-$722 showed rather weak sign of pulsations 
as shown in Figure \ref{fig:psd} (b). 

\placefigure{fig:psd}

For the 11 sources from which pulsations were detected in the FFT analysis,
we performed an epoch folding search to determine
more precise pulse periods.
In addition,
the orbital Doppler effect was corrected for
SMC X-1, using the ephemeris presented in Wojdowski et al. (1998)\markcite{Wojdowski1998}.
These derived pulse periods are presented in Table \ref{tab:pcat}.

\placetable{tab:pcat}

\begin{figure}
\psbox[xsize=0.47\textwidth]{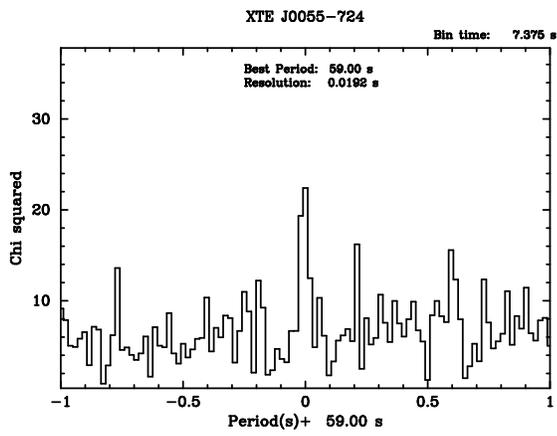}
\caption{Periodigram of XTE J0055$-$724 around a known period $\sim 59$~s.
\label{fig:efs}
}
\end{figure}

We found no coherent period in the FFT power spectra
from three sources positionally coincident with known pulsars: 
XTE J0055$-$724 (No.\,16), SMC X-1 (No.\,38; in observation C),
and RX J0052.1$-$7319 (No.\,14).
Nevertheless, we tried the epoch folding search\footnote{Since
SMC X-1 was in the 0.6-day eclipse phase during observation C,
we only used the data from non-eclipse times. }. 
A weak peak was detected from source No.\,16 
near the known period $\sim 59$~s as shown in Figure \ref{fig:efs}, 
which indicates that No.\,16 is a counterpart of XTE J0055$-$724. 
This period is, however, not referred in Table \ref{tab:pcat}, 
simply because this source was relatively dim, 
and any reliable period was not obtained.

\placefigure{fig:efs}

 We also searched for burst-like activity
by using light curves binned with various time scales
from $\sim$1~second to $\sim$1~hour.
However, no bursts were found.

\subsection{Spectral Analysis \label{sec:spec}}
 X-rays were detected from eight radio SNRs:
0102$-$723 (No.\,30), 0103$-$726 (No.\,31), N19 (No.\,1), N66 (No.\,24),
DEM S128
(No.\,32\footnote{No.\,32=DEM S128 (radio SNR)=AX J0105$-$722 (X-ray pulsar).}),
0056$-$725 (No.\,21), 0049$-$736 (No.\,12), and
0047$-$735 (No.\,4\footnote{Excluded from the spectral analysis
because of contamination (see \S\ref{subsec:image}).}).
Before fitting the overall spectra,
we distinguished thermal SNRs from others.
Direct evidence for a thermal spectrum is the presence of
emission lines from highly ionized atoms.
The brightest SNR 0102$-$723 is already known to exhibit thermal X-ray
emission
(e.g. Hayashi et al. 1994\markcite{Hayashi1994}), hence we examined whether or not
the spectra of the remaining SNRs show any emission lines.
 In order to obtain better energy resolution,
we extracted the SIS spectra of 0103$-$726, N19, and N66.
The other SNRs DEM S128, 0056$-$725, and 0049$-$736 were out of the SIS
fields.
Since the latter two were too faint for spectral fitting,
we only used the GIS spectrum of DEM S128.
 We fitted the spectra in a narrow energy band
($\sim 1.1$--3~keV)
with a bremsstrahlung continuum and three narrow Gaussian lines centered
at 1.34~keV, 1.85~keV, and 2.46~keV
(K$_\alpha$ lines from He-like Mg, Si, and S), and we
found evidence of emission lines from three SNRs
(0103$-$726, N19, and N66).
We thus regarded them as thermal SNRs,
and fitted their SIS spectra with
thin-thermal plasma models as described in \S\ref{subsec:snrs}.
For the other SNRs,
we adopted both a power-law model and
a thin-thermal plasma model (Raymond \& Smith 1977\markcite{Raymond1977}),
with interstellar absorption.
Since their GIS spectra\footnote{Only the GIS2 spectrum was used for 0049$-$736,
because this SNR was detected in the proximity of the GIS3
calibration source.}
had poor photon statistics,
we fixed the global abundance to be
0.2~solar,
i.e. that of the interstellar matter in the SMC
(Russell \& Dopita 1992\markcite{Russell1992}),
when the thermal plasma model was applied.
We hereafter refer to this abundance value as ``the SMC abundance.''

 As summarized in Table \ref{tab:pcat}, we detected 13 X-ray pulsars.
In addition, No.\,22 and No.\,13 are
positionally coincident with
RX J0058.2$-$7231 and RX J0051.9$-$7311,
both are HMXBs with a Be star companion
(Cowley et al. 1997\markcite{Cowley1997}; 
Schmidtke et al. 1999\markcite{Schmidtke1999}).
X-ray spectra of these classes are generally described by
an absorbed power-law below $\sim 10$~keV (e.g. Nagase 1989\markcite{Nagase1989}) and
hence we used this model for spectral fitting.
Spectra of the other (unclassified) sources had poor statistics,
hence it was not clear whether they are thermal or not.
Nevertheless, we also adopted the power-law model for them.
We used only GIS2 spectra of
AX J0051$-$733 (No.\,8),
RX J0059.2$-$7138 (No.\,25),
SMC X-1 (in observation H),
and RX J0051.9$-$7311,
because RX J0059.2$-$7138
was detected only on the edge of GIS2 and
the other three were detected near the GIS3 calibration source.
On the other hand,
only the GIS3 spectrum of
AX J0049$-$729 (No.\,3; in observation G)
was used because
it was detected near the calibration source of GIS2.
In other cases we added the spectra of GIS2 and GIS3 to obtain better statistics.

 The simple power-law model showed a soft excess for three sources:
RX J0059.2$-$7138, XTE J0111.2$-$7317 (No.\,37)
and SMC X-1 (in each observation).
Including blackbody emission as a soft component gave a better fit.
Spectral parameters,
fluxes and absorption-corrected luminosities in Table \ref{tab:smccat}
are derived from this two-component model for these sources.

 Model fitting for the spectra of the SNR 0047$-$735 (No.\,4),
the X-ray pulsar AX J0049$-$732 (No.\,5),
and No.\,34 was not performed because these are heavily contaminated
by nearby sources (see \S\ref{subsec:cat}).
Sources No.\,27, No.\,28, and 1SAX J0103.2$-$7209 (No.\,29)
have been observed more than once, and the
spectral parameters were derived separately.
Since the parameters are consistent between observations
for all three sources, we performed combined fitting
for the spectra of different observations.
However,
AX J0049$-$729 and
SMC X-1, which have also been observed more than once,
showed large flux variability and different spectral parameters.
 The SNR 0102$-$723 (No.\,30) has been detected twice in the GIS field
(in observation B and D), but only once in the SIS field (B).
Since this source is a line-dominated young SNR, we only used the SIS spectrum
for the spectral analysis (see \S\ref{subsec:snrs}).

 The derived spectral parameters: photon index $\Gamma$ or temperature $kT$,
and column density $N_{\rm H}$, flux $F_{\rm x}$,
and absorption-corrected luminosity $L_{\rm x}$
are presented in Table \ref{tab:smccat}.

\section{Comments on Specific Sources \label{sec:comments}}

\subsection{SNRs \label{subsec:snrs}}

\begin{figure}
\psbox[xsize=0.47\textwidth]{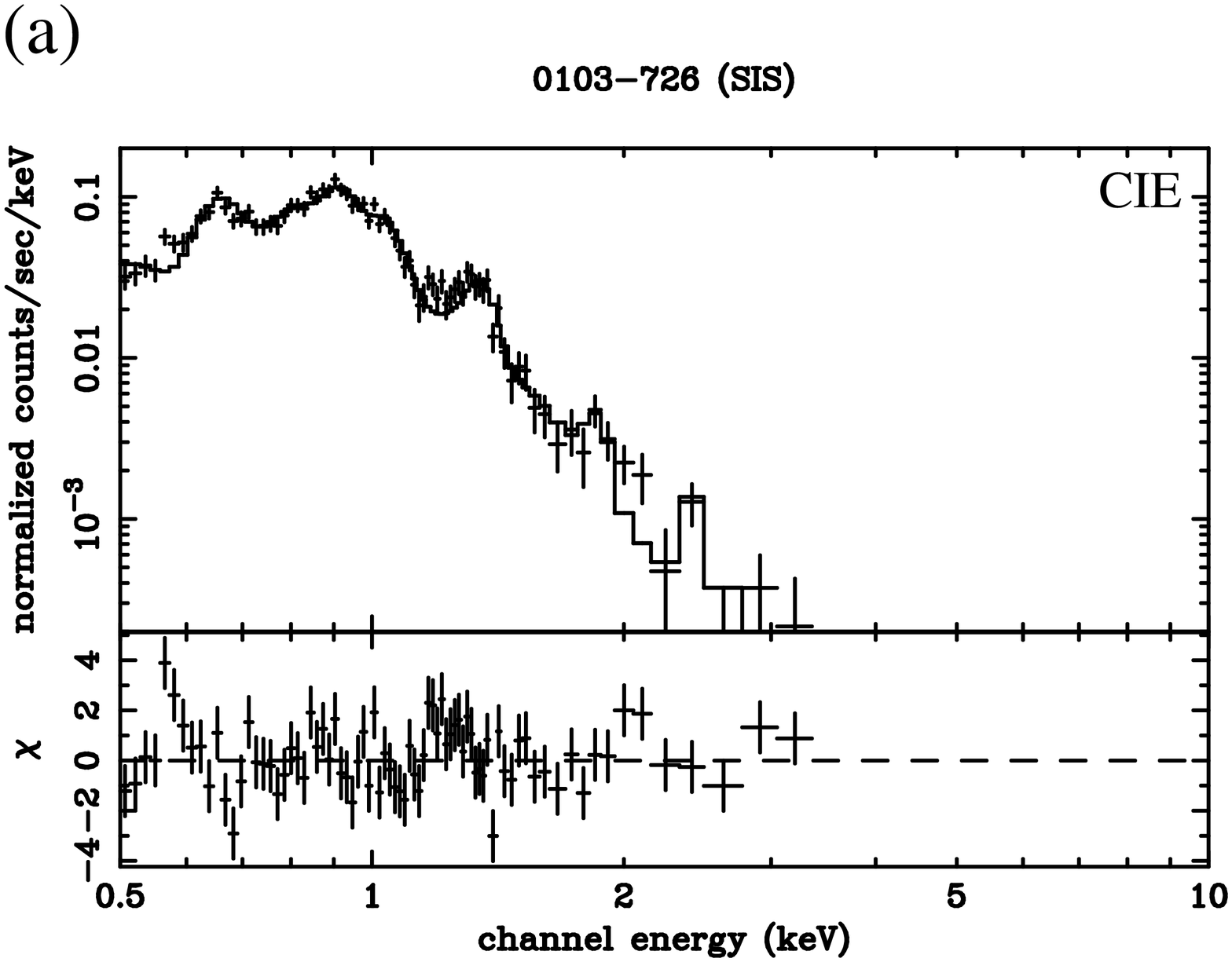}
\psbox[xsize=0.47\textwidth]{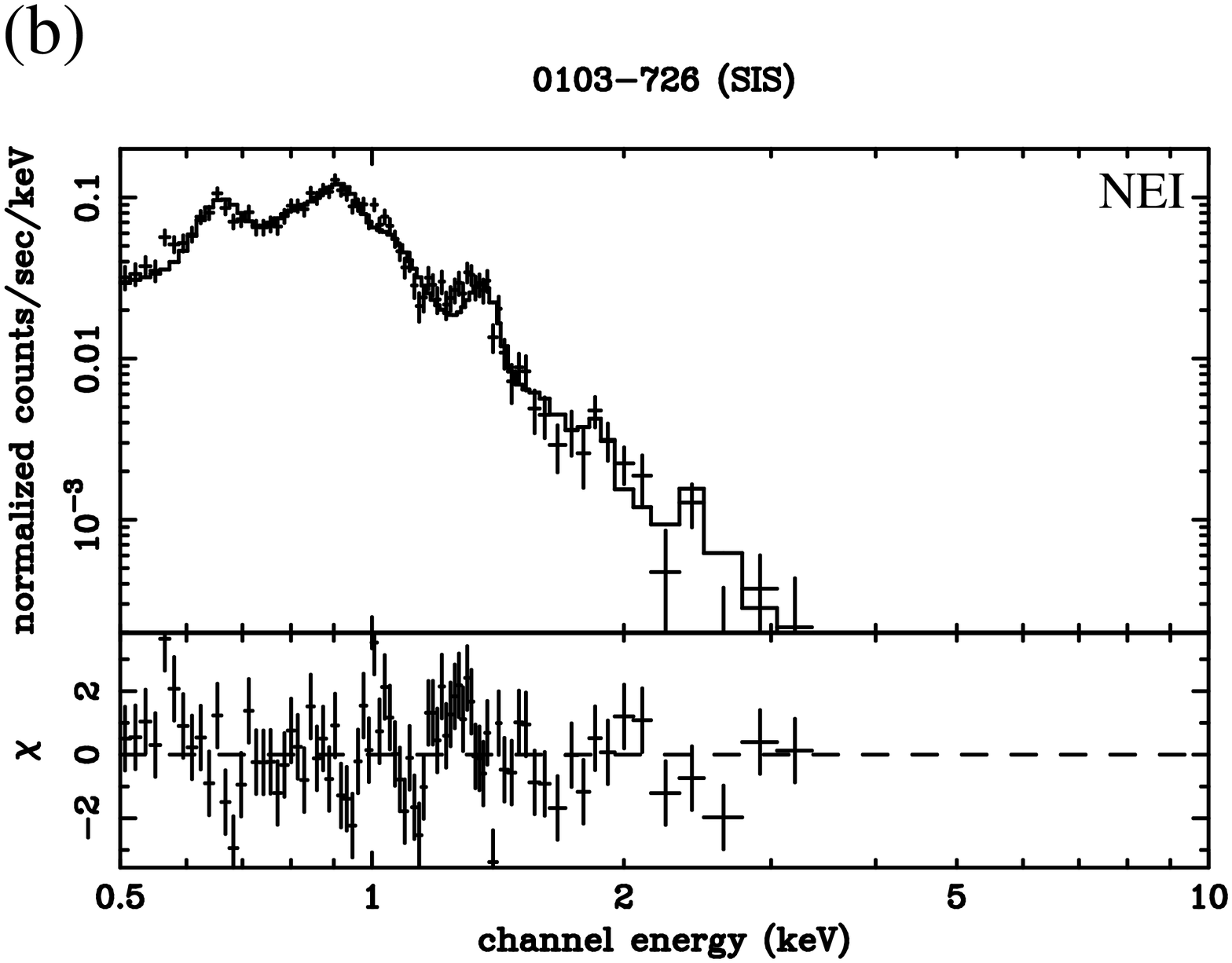}
\psbox[xsize=0.47\textwidth]{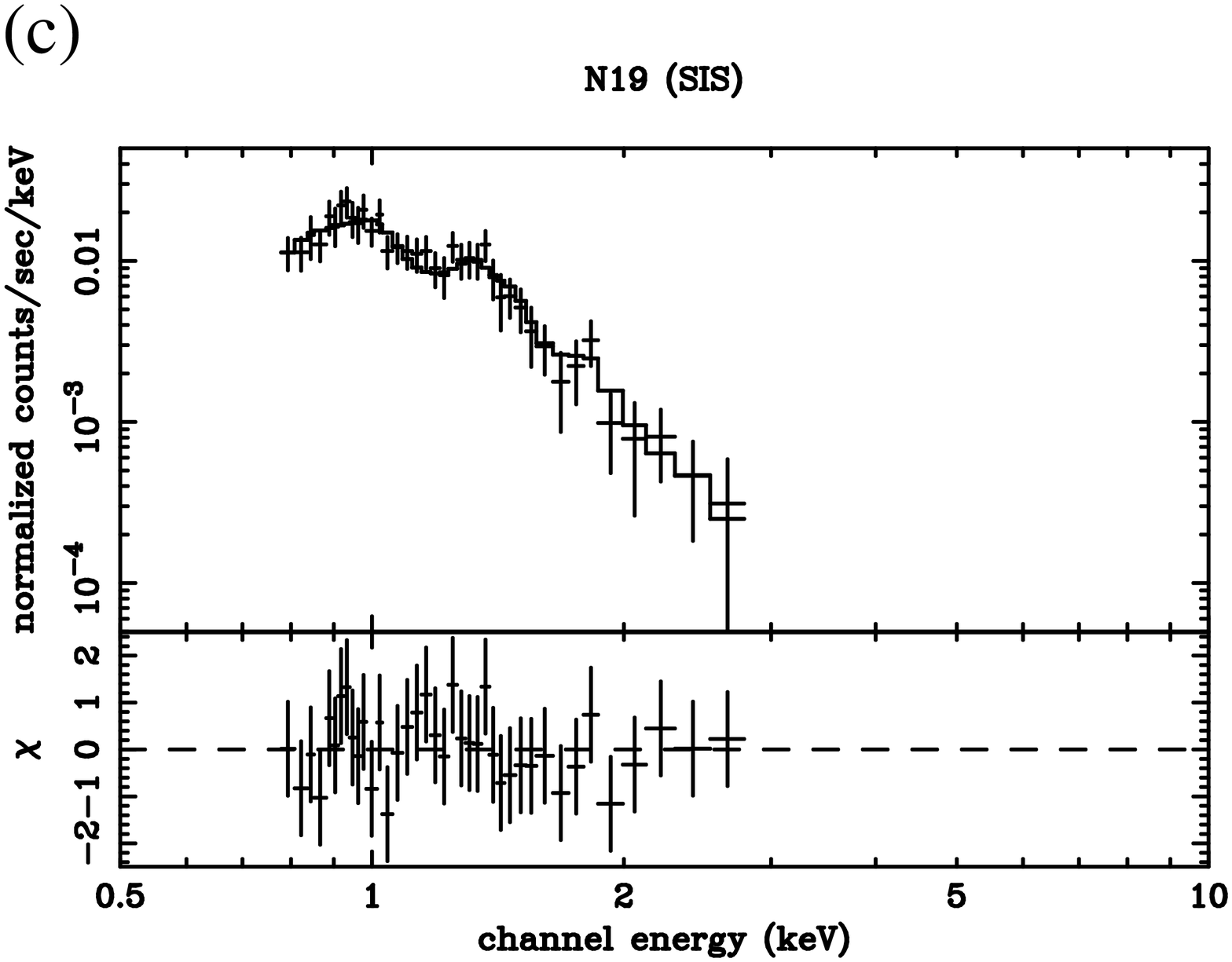}
\psbox[xsize=0.47\textwidth]{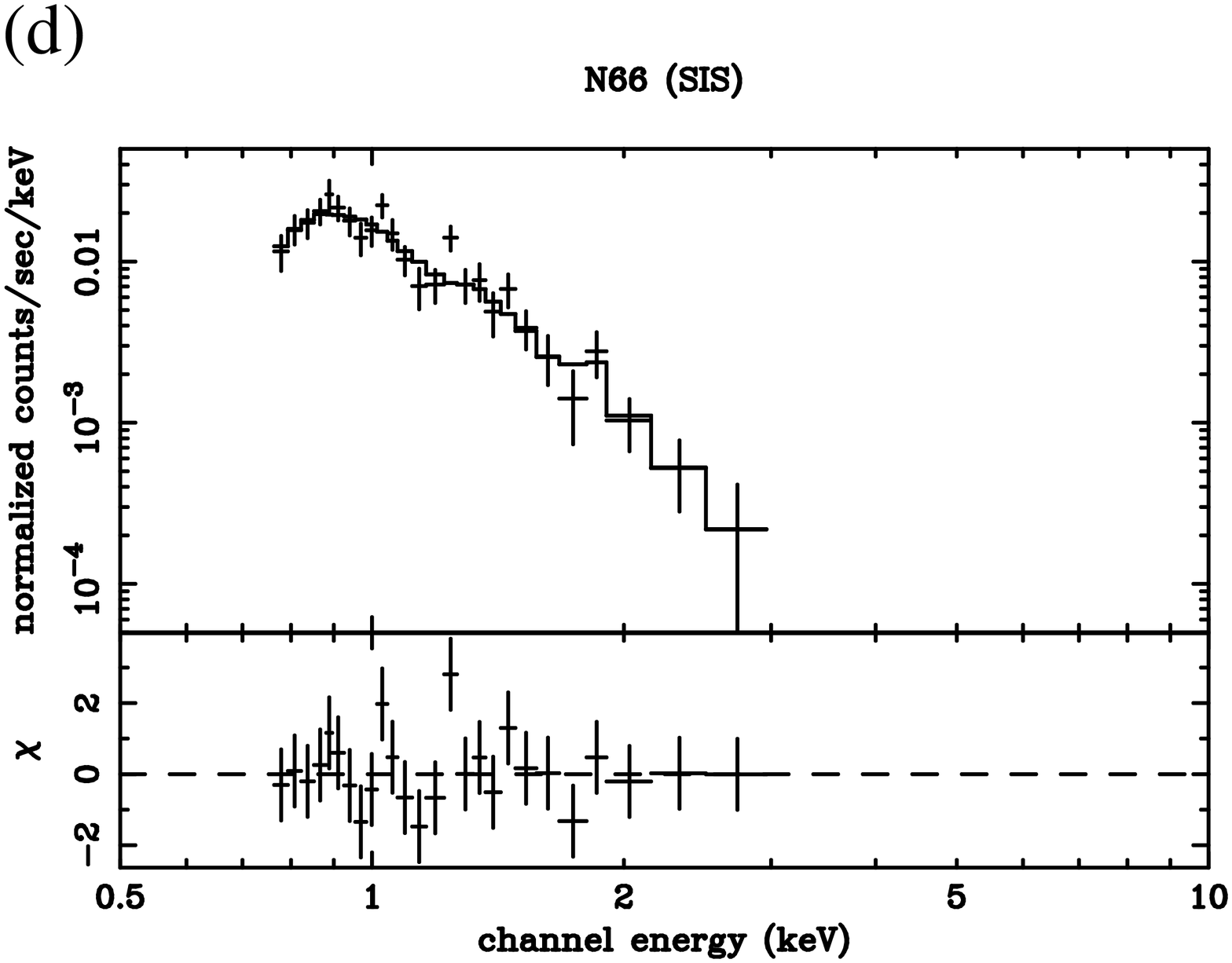}
\psbox[xsize=0.47\textwidth]{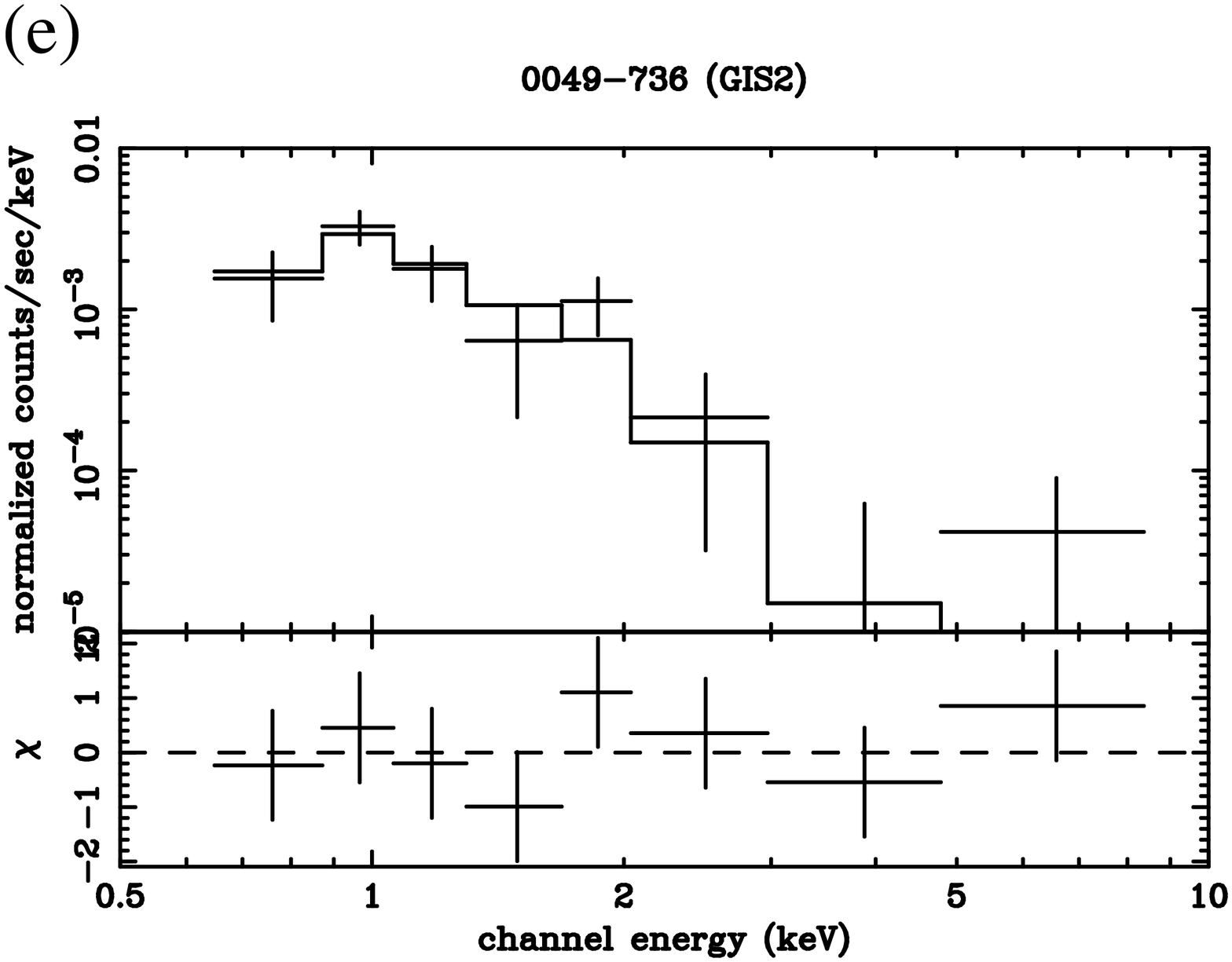}
\psbox[xsize=0.47\textwidth]{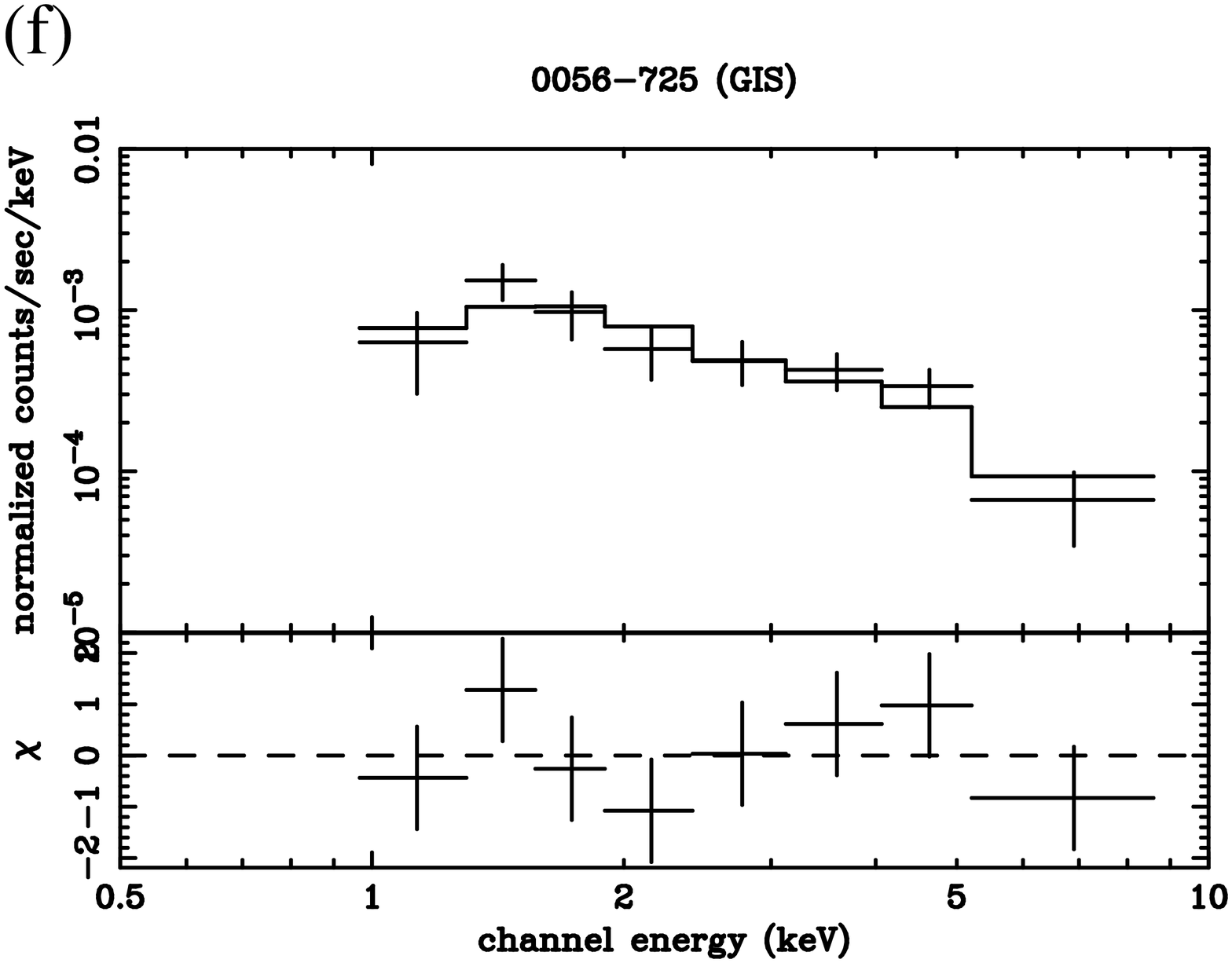}
\psbox[xsize=0.47\textwidth]{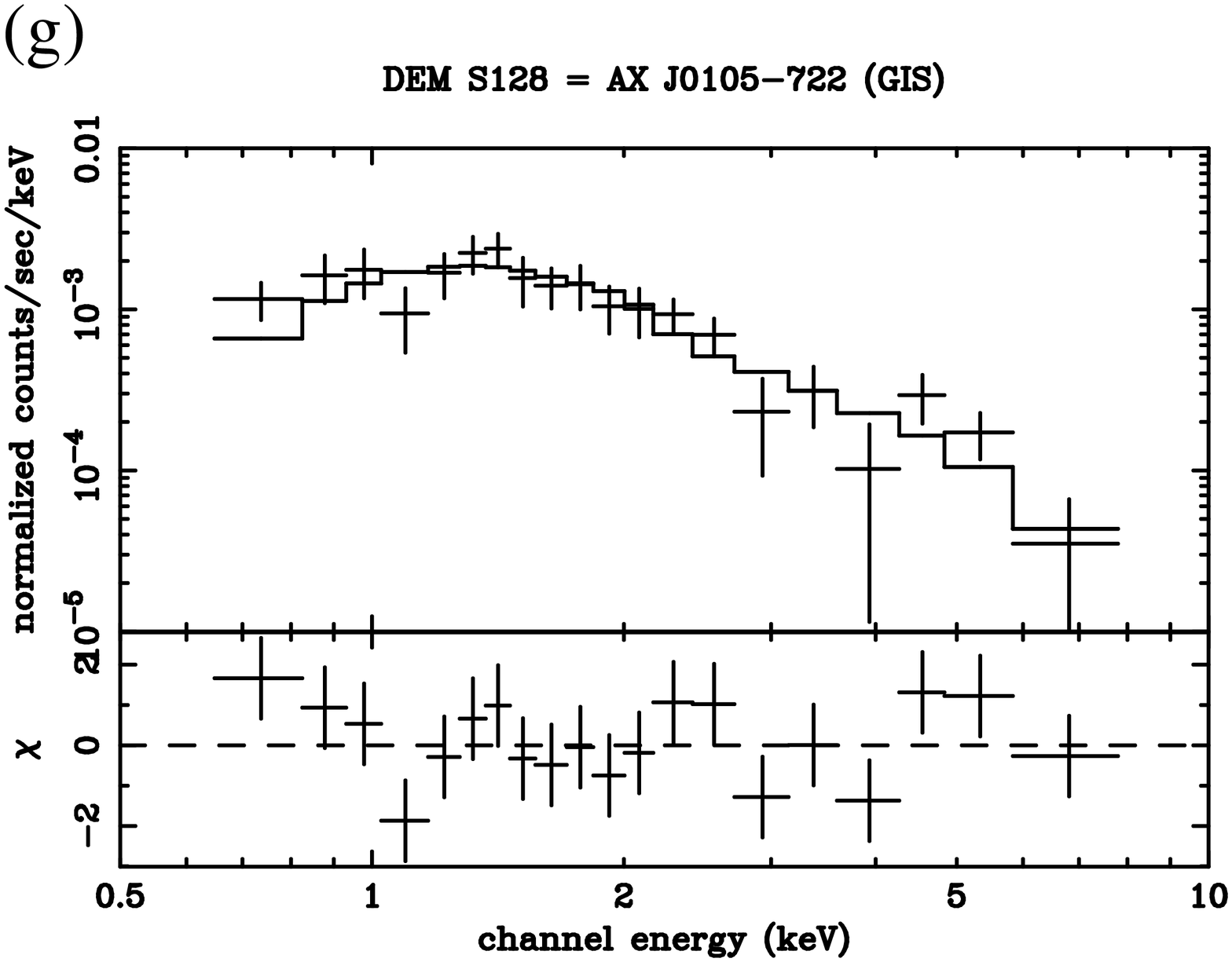}
\caption{Spectra of SNRs. 
The upper panels show data points (crosses) and the best-fit model 
(solid line; see text), 
while the lower panels show residuals from the best-fit model. 
(a) 0103$-$726 with the CIE model; 
(b) 0103$-$726 with the NEI model; 
(c) N19; 
(d) N66; 
(e) 0049$-$736; 
(f) 0056$-$725; 
(g) DEM S128 (=AX J0105$-$722).  
\label{fig:specsnrs}
}
\end{figure}

Figure \ref{fig:specsnrs} gives the spectra of SNRs
with the best-fit model described below.
The relevant parameters are presented in Table
\ref{tab:smccat} and \ref{tab:spec0103}.

\paragraph{No.\,30: 0102$-$723}

A detailed analysis of the SIS spectrum
of this SNR was carried out by Hayashi et al. (1994)\markcite{Hayashi1994}.
We adopted their model and determined
its flux and absorption-corrected luminosity
as in Table \ref{tab:smccat}.

\paragraph{No.\,31: 0103$-$726}

We first adopted an absorbed thin-thermal plasma model
(Raymond \& Smith 1977\markcite{Raymond1977}), in which
the condition of collisional ionization equilibrium (CIE)
is assumed.
The abundances of O, Ne, Mg, Si, and S were treated as free parameters,
while those of the other elements were fixed to the SMC abundance.
The best-fit parameters and
the model spectrum are given in
Table \ref{tab:spec0103} and Figure \ref{fig:specsnrs} (a), respectively.
Although the CIE model could reproduce the line profile fairly well,
it was statistically rejected with 99\% confidence.

\begin{figure}
\psbox[xsize=0.47\textwidth]{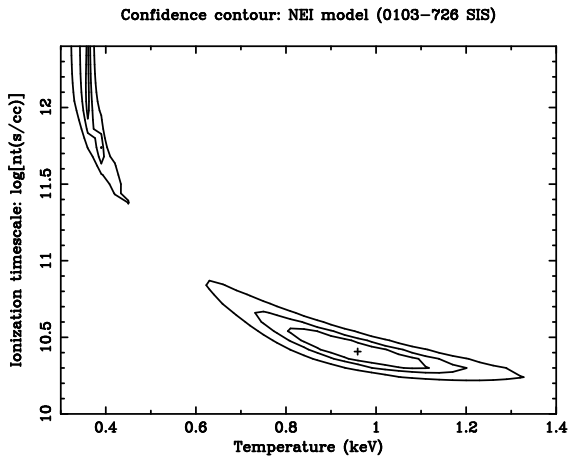}
\caption{Confidence contours between $\tau$ and $kT$ 
in the NEI model for 0103$-$726. 
\label{fig:0103contour}
}
\end{figure}

We then included the effects of non-equilibrium ionization (NEI)
in the plasma model (Masai 1994\markcite{Masai1994}).
An additional parameter is the ionization timescale
$\tau = nt $, where $n$ is the electron density and $t$
is the elapsed time after the plasma has been heated.
Note that the plasma is in the CIE condition when
$\tau$ is larger than $\sim 10^{12}$~s~cm$^{-3}$.
The best-fit parameters and
the model spectrum are given in
Table \ref{tab:spec0103} and Figure \ref{fig:specsnrs} (b), respectively.
This model was also statistically rejected with 99\% confidence.
The larger $\chi^2$ is due to the difference between
the CIE/NEI plasma codes.
The best-fit value of the ionization timescale
indicates that the plasma of 0103$-$726 is in the NEI condition.
However, the confidence contours shown in Figure \ref{fig:0103contour} 
suggest that we can not distinguish
between a high temperature NEI plasma
and a low temperature CIE plasma model.

\placetable{tab:spec0103}

\paragraph{No.\,1: N19}

We adopted an absorbed thin-thermal CIE plasma model.
We set the abundances of Ne and Mg to be free,
while those of the other elements were fixed to the SMC abundance.
Best-fit values of the abundances are
0.7 (0.3--1.8) for Ne and 1.0 (0.3--1.8) for Mg
(hereafter, 90\% confidence limits are given in parentheses,
unless otherwise mentioned).

\paragraph{No.\,24: N66}

An absorbed thin-thermal CIE plasma model,
where the SMC abundance was assumed for all elements,
could well reproduce the SIS spectrum.

\paragraph{No.\,12: 0049$-$736}

Although both a thermal CIE model and
a power-law model could well describe the spectrum,
we adopted the thermal one
because the derived temperature ($\sim 0.7$~keV)
is reasonable for an SNR.
This implies that 0049$-$736 may be a thermal SNR.

\paragraph{No.\,21: 0056$-$725}
Since a thermal CIE model yielded
an unusually high temperature of $\sim 20$ ($>4$)~keV,
we adopted a power-law model.
It may be implied that
the spectrum of 0056$-$725 really has a non-thermal origin.

\paragraph{No.\,32: DEM S128}
A thermal CIE model gave
a temperature of 3.2 (2.2--5.0)~keV,
which may be reasonable if this SNR is very young.
However, generally
such a young SNR exhibits prominent emission lines
which originate from shock-heated ejecta.
The fact that no emission line was identified
prefers the power-law model,
which is adopted in Figure \ref{fig:specsnrs} (g) and Table \ref{tab:smccat}.
Therefore a non-thermal spectrum for DEM S128 is implied.
In addition, we have
detected evidence for coherent pulsations
from this source (= AX J0105$-$722)
as described in \S\ref{sec:timing}.
The nature of DEM S128 is discussed in \S\ref{subsec:dems128}.

\placefigure{fig:specsnrs}
\placefigure{fig:0103contour}

\subsection{X-ray Pulsars \label{sec:pul}}

\begin{figure}
\psbox[xsize=0.47\textwidth]{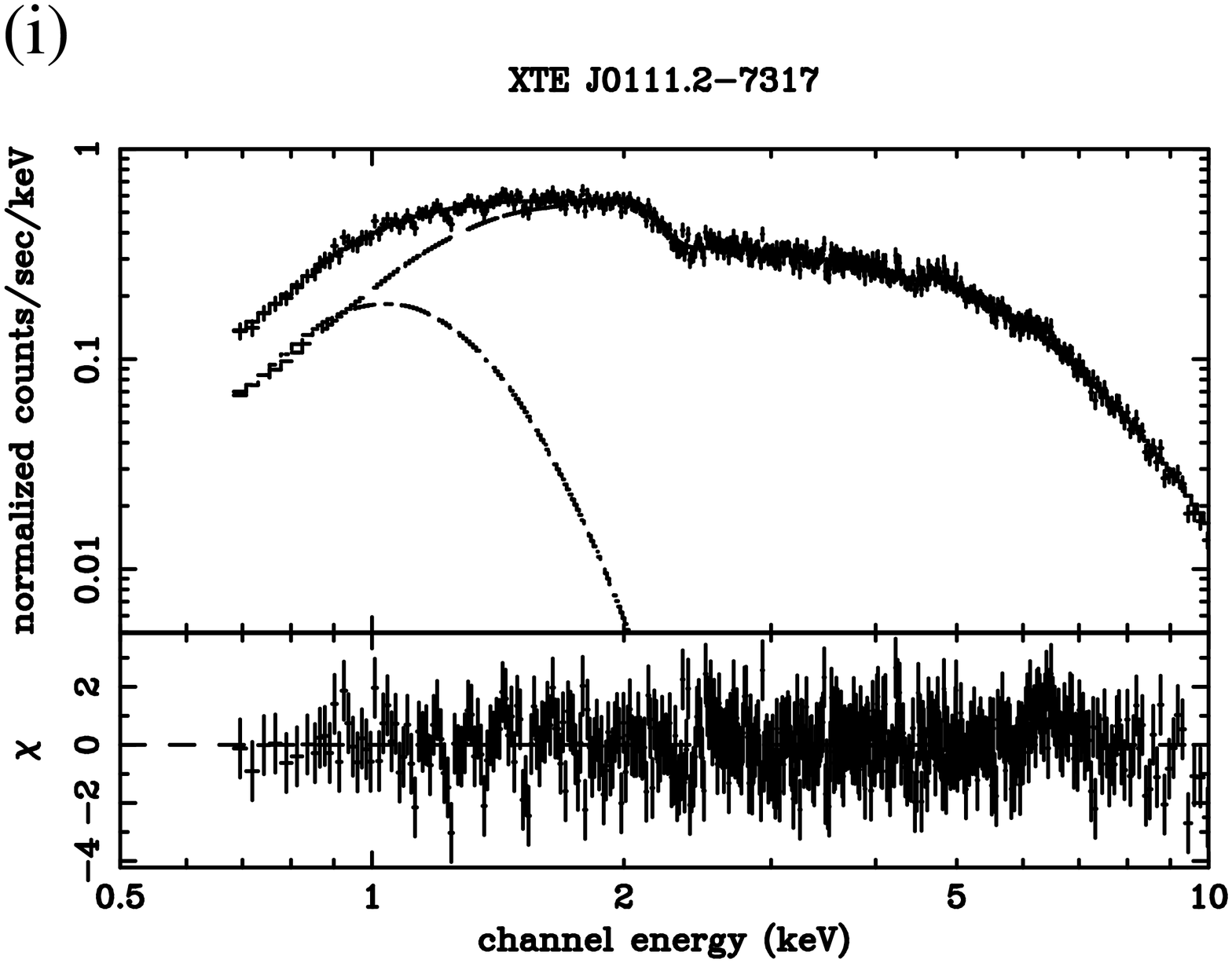}
\psbox[xsize=0.47\textwidth]{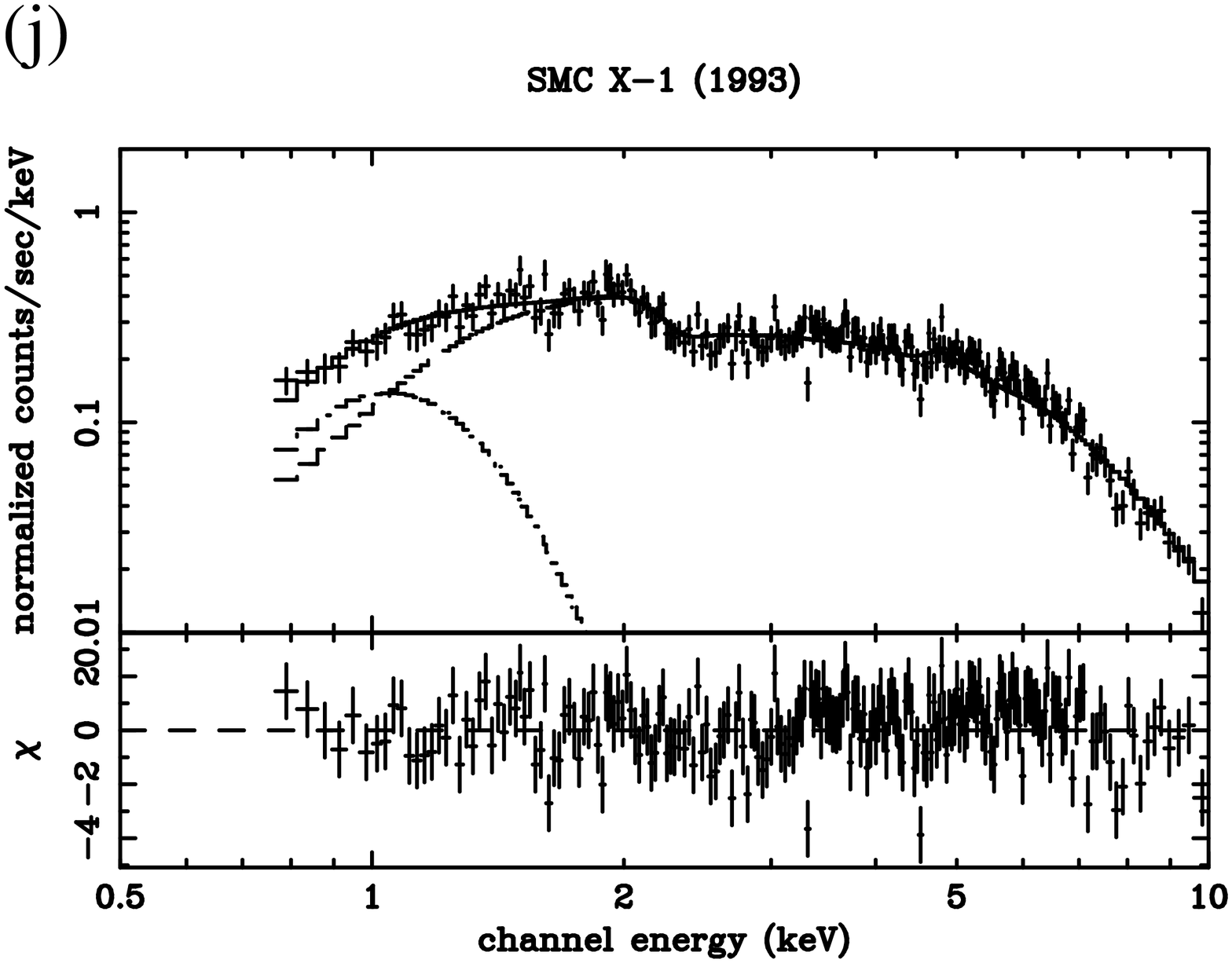}

\psbox[xsize=0.47\textwidth]{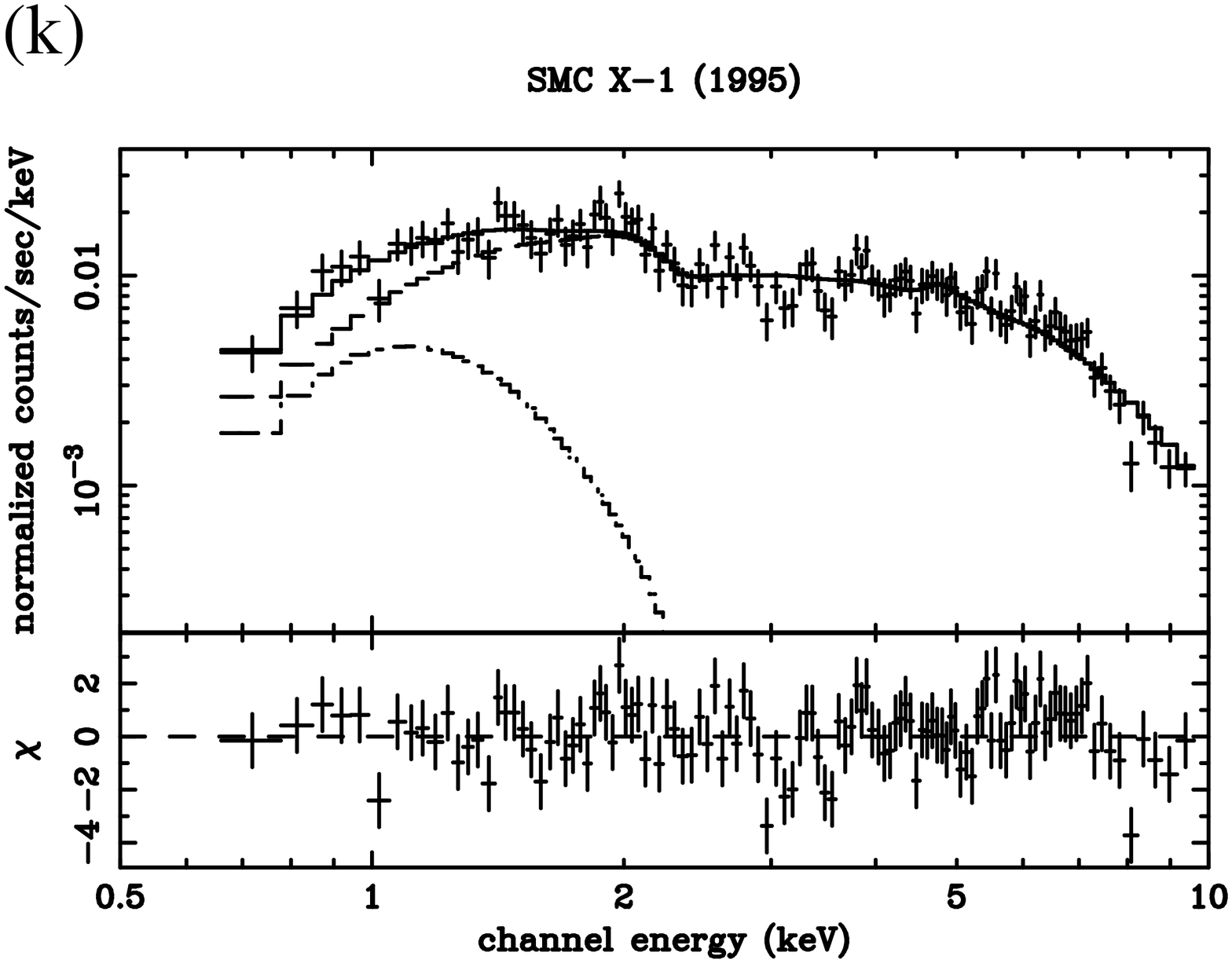}
\psbox[xsize=0.47\textwidth]{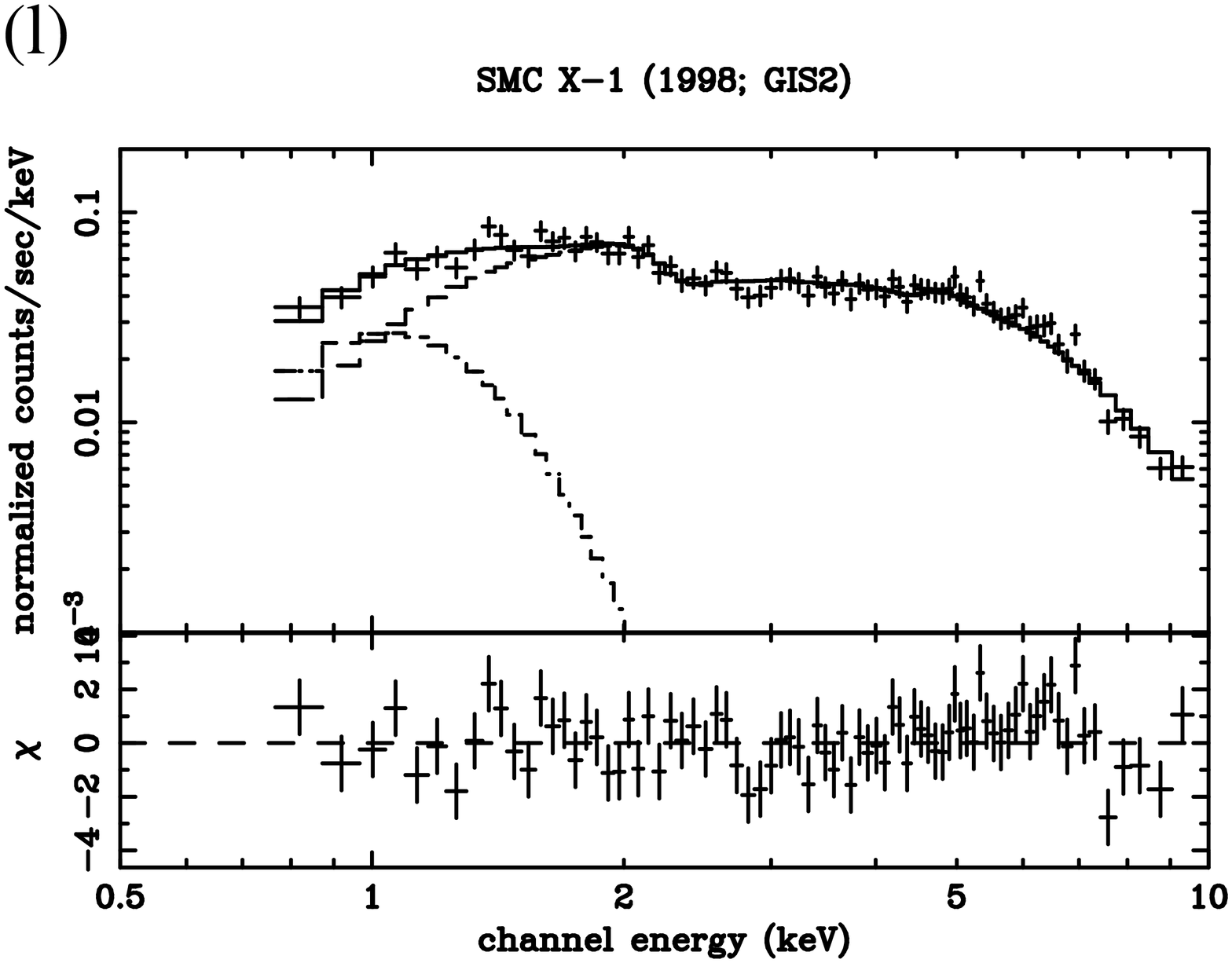}
\caption{GIS spectra of X-ray pulsars. 
The upper panels show data points (crosses) and the best-fit model 
(solid line; see text), 
while the lower panels show residuals from the best-fit model. 
See Figure 4 (g) for AX J0105$-$722. 
(a) AX J0049$-$729 in observation E; 
(b) AX J0051$-$733; 
(c) AX J0051$-$722;
(d) 1WGA J0053.8$-$7226;
(e) XTE J0055$-$724; 
(f) AX J0058$-$7203; 
(g) RX J0059.2$-$7138; 
(h) 1SAX J0103.2$-$7209 (average of observations B, D, and F); 
(i) XTE J0111.2$-$7317; 
(j) SMC X-1 in observation A; 
(k) SMC X-1 in observation C; 
(l) SMC X-1 in observation H. 
\label{fig:pulspec}
}
\end{figure}

\begin{figure}
\psbox[xsize=0.47\textwidth]{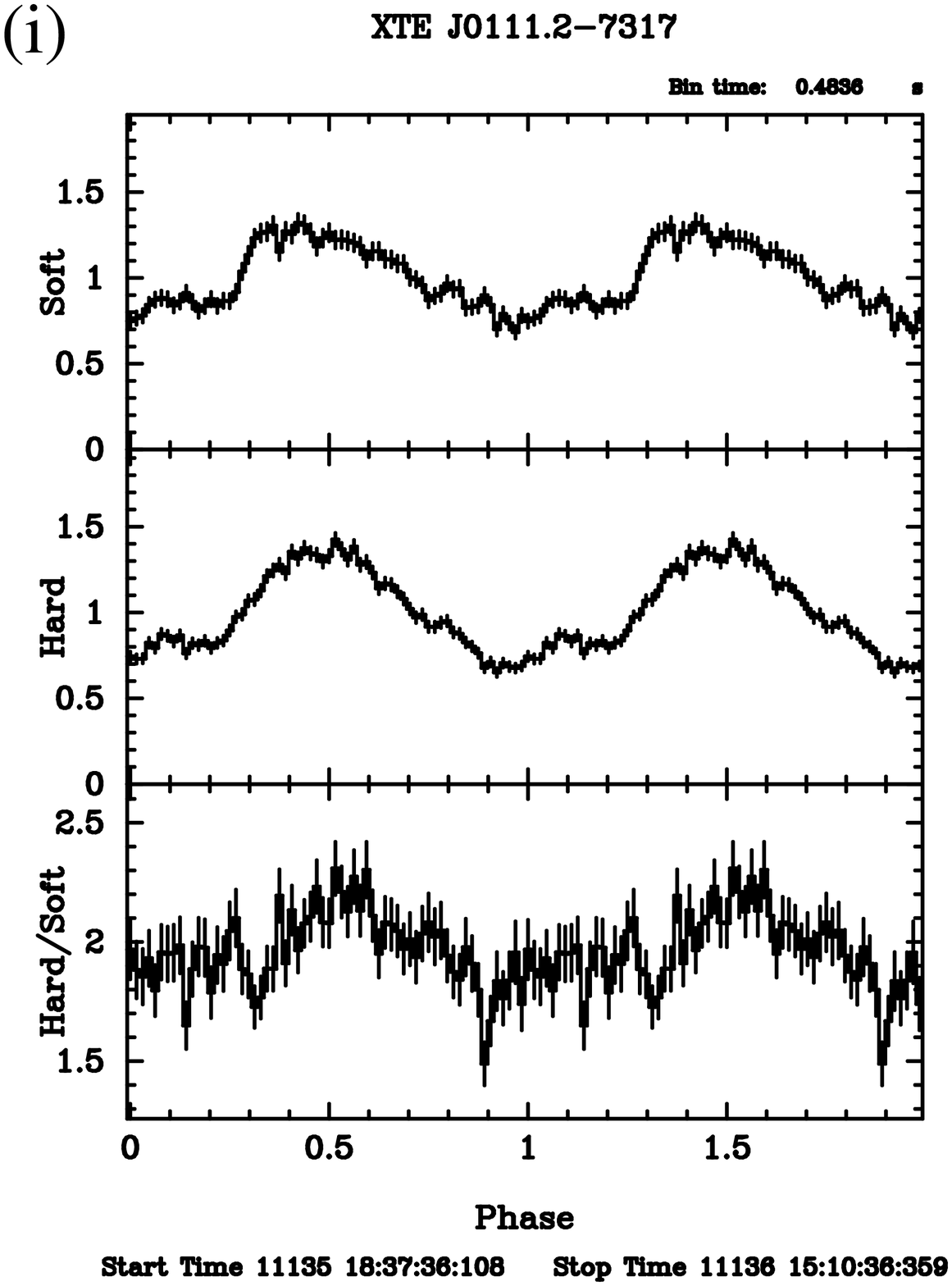}
\psbox[xsize=0.47\textwidth]{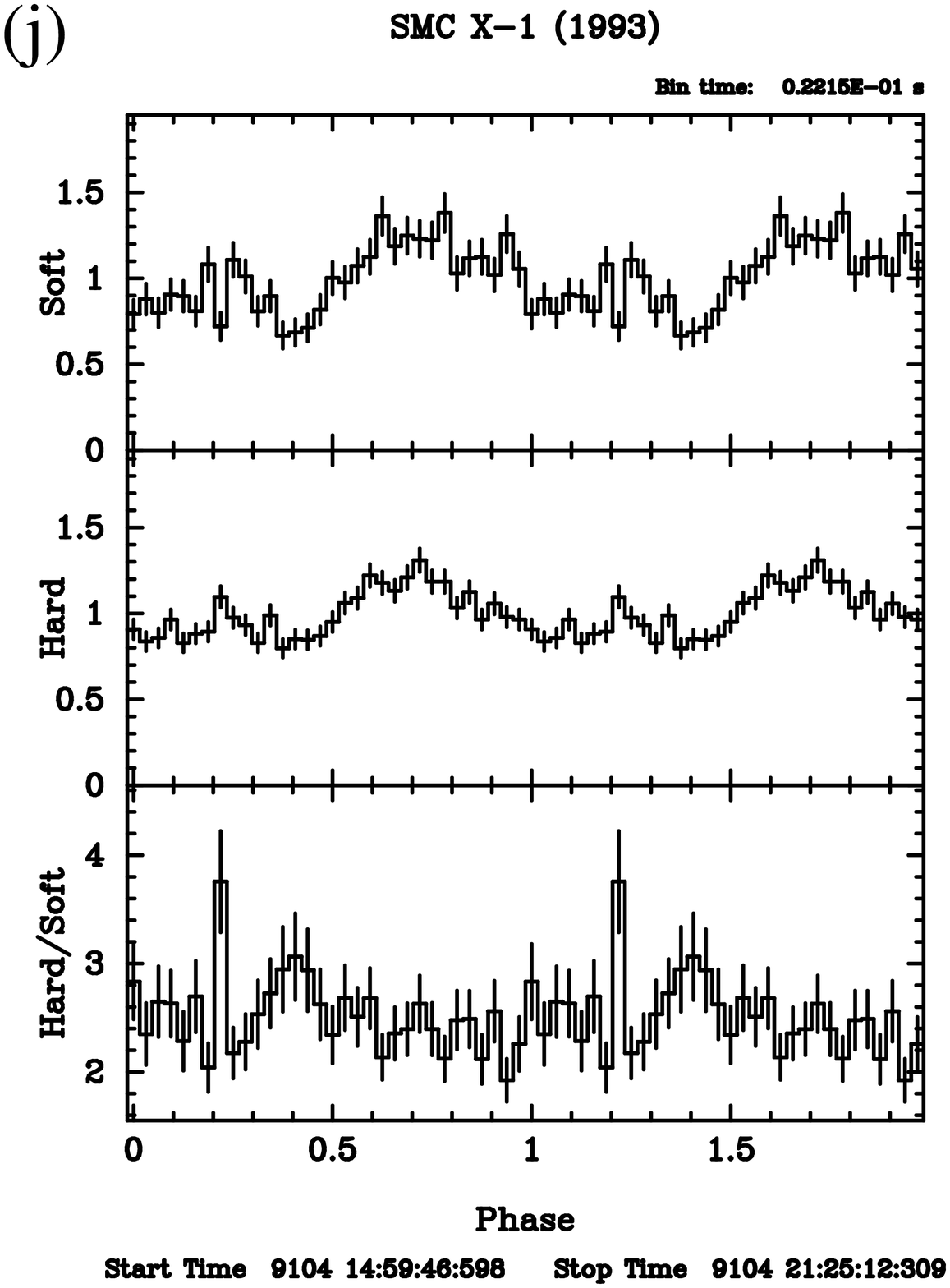}

\psbox[xsize=0.47\textwidth]{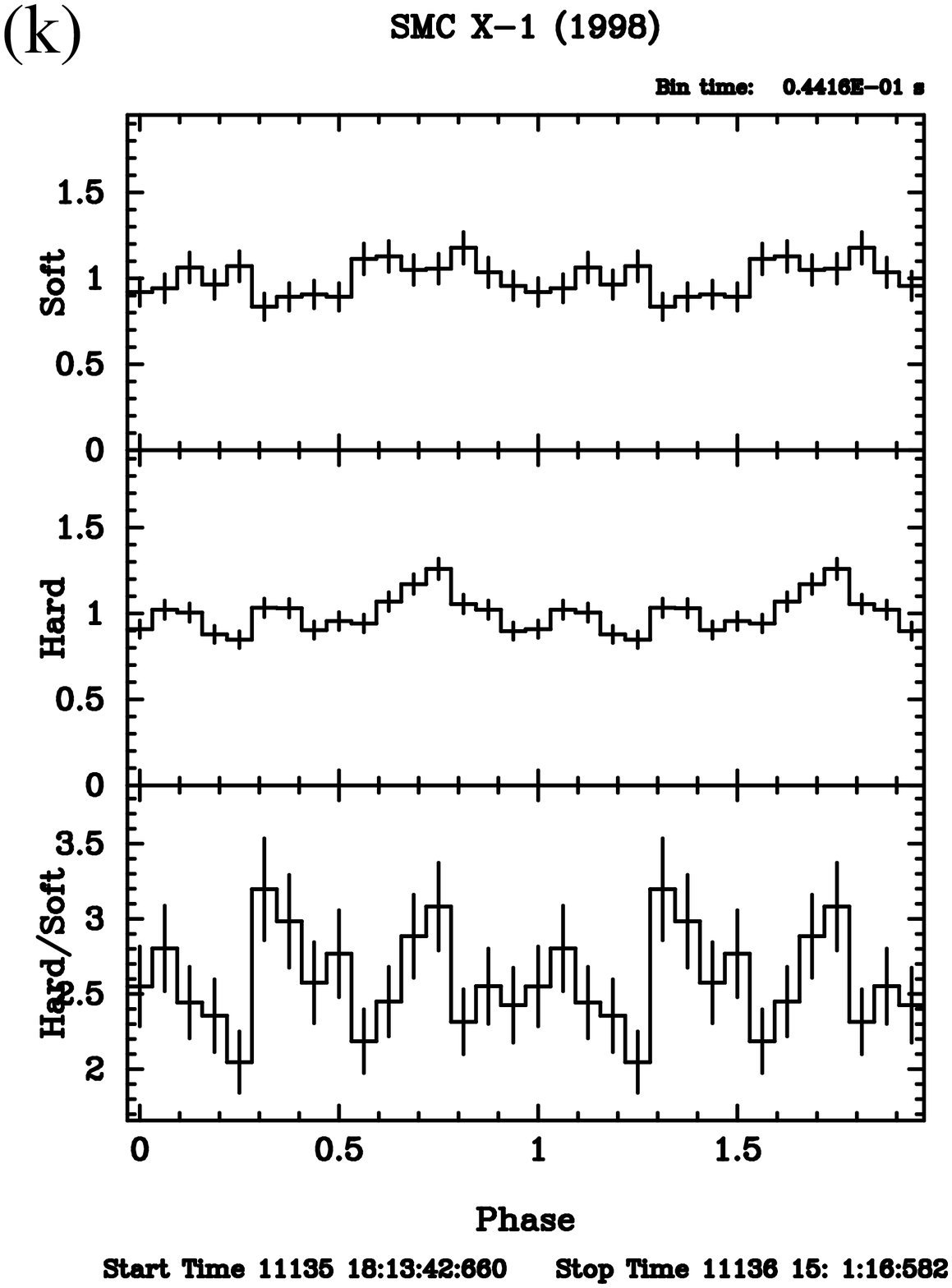}
\caption{Folded light curves of X-ray pulsars in the
soft (0.7--2.0~keV; upper panel) and
hard (2.0--7.0~keV; middle panel) band, in which 
the vertical axes show the normalized count rates.
The lower panel shows the intensity ratio of
the hard band to the soft band.
(a) AX J0049$-$729 in observation E; 
(b) AX J0051$-$733; 
(c) AX J0051$-$722;
(d) 1WGA J0053.8$-$7226;
(e) AX J0058$-$7203; 
(f) RX J0059.2$-$7138; 
(g) 1SAX J0103.2$-$7209 in observation D; 
(h) AX J0105$-$722; 
(i) XTE J0111.2$-$7317; 
(j) SMC X-1 in observation A; 
(k) SMC X-1 in observation H. 
\label{fig:pullc}
}
\end{figure}

 In this subsection, pulse periods are presented with
errors
of the last digits
in parentheses. The spectra and energy-resolved pulse shapes
are separately
presented in Figures \ref{fig:pulspec} and \ref{fig:pullc}, respectively.
We suggest that 14 of the 16 pulsars in the SMC
are X-ray binary pulsars (XBPs),
because of their flux variability and,
in some cases, existence of an optical counterpart.

\paragraph{No.\,3: AX J0049$-$729}
Coherent pulsations with a 74.8(4)~s period
were first detected with RXTE (Corbet et al. 1997a\markcite{Corbet1997a})
in the direction of SMC X-3.
However, the positional uncertainty was very large ($\sim 2^\circ$).
During this analysis,
we found that No.\,3 (AX J0049$-$729) was pulsating with a 74.68(2)~s period
and so determined its position
more accurately
(Yokogawa \& Koyama 1998a\markcite{Yokogawa1998a}; 
Yokogawa et al. 1999a\markcite{Yokogawa1999a}).
The ROSAT counterpart of this pulsar, RX J0049.1$-$7250,
provides the most accurate position with a $\pm 13''$ error circle
(Kahabka \& Pietsch 1998\markcite{Kahabka1998}),
in which one Be star has been discovered (Stevens et al. 1999\markcite{Stevens1999}).

This pulsar has been included in 11 observation fields
of Einstein, ROSAT, and ASCA.
Yokogawa et al. (1999a\markcite{Yokogawa1999a})
found
a large flux variability by a factor $\gtrsim 100$
during the 11 observations,
thus we can conclude that
AX J0049$-$729 is an XBP with a Be star companion.

\paragraph{No.\,5: AX J0049$-$732}
Coherent pulsations with a 9.1321(4)~s period
from No.\,5 (AX J0049$-$732) were first detected during this analysis
(Imanishi, Yokogawa, \& Koyama 1998\markcite{Imanishi1998}),
from the data in a circle with a 3$^\prime$ radius
centered on AX J0049$-$732.
However, as noted in \S\ref{subsec:cat},
the separation between AX J0049$-$732
and source No.\,4 (SNR 0047$-$735) is only $\sim 2^\prime$.
 Therefore, we also performed an FFT analysis on
the data in a circle with 3$^\prime$ radius centered on the SNR,
and detected no pulsations.
 Hence the pulsations can clearly be attributed
to AX J0049$-$732.

ROSAT sources 1WGA J0049.4$-$7310
and 1WGA J0049.1$-$7311 are in the error circle of AX J0049$-$732.
We took their ROSAT spectra from HEASARC archive system
and fitted each of them with an absorbed power-law model.
The sum of their fluxes was determined to be
$\sim 4 \times 10^{-14}$\uflux\ (0.3--2.0~keV).
As noted in \S\ref{subsec:cat},
the ASCA spectrum of AX J0049$-$732 was heavily contaminated
by that of a soft source No.\,4 (SNR 0047$-$735). 
Hence an accurate estimation of its flux was difficult,
especially in the soft band.
Nevertheless,
we fitted the contaminated spectrum simply with
an absorbed power-law model and
the flux was derived to be $\sim 9 \times 10^{-14}$\uflux\ (0.3--2.0~keV).
Further observations with
better spatial resolution and photon statistics are needed
to reveal the nature of AX J0049$-$732.

\paragraph{No.\,8: AX J0051$-$733}
Coherent pulsations with a 323.2(5)~s period
from No.\,8 (AX J0051$-$733) were first detected during this analysis
(Yokogawa \& Koyama 1998b\markcite{Yokogawa1998b}).
RX J0050.8$-$7316, the ROSAT HRI counterpart of AX J0051$-$733,
has a Be star in its error circle (Cowley et al. 1997\markcite{Cowley1997}).
In addition, Imanishi et al. (1999)\markcite{Imanishi1999}
investigated all the 
observation fields of Einstein and ROSAT that included AX J0051$-$733, and
found flux variations of a factor $\gtrsim 10$.
These results indicate that
AX J0051$-$733 is an XBP with a Be star companion.

\paragraph{No.\,11: AX J0051$-$722}
Coherent pulsations with a $\sim92$~s period were first discovered
with an RXTE observation
on November 15, 1997 (Marshall et al. 1997\markcite{Marshall1997}).
A TOO (Target Of Opportunity) observation with ASCA on December 12 (observation G)
revealed that No.\,11 was pulsating with 91.12(5)~s period
(Corbet et al. 1997a\markcite{Corbet1997a}).
Since then,
this source has been detected in many RXTE/BeppoSAX/ASCA observations,
and found to be fading and rebrightening
(Lochner et al. 1998\markcite{Lochner1998a};
Lochner 1998\markcite{Lochner1998b}).
Therefore it is certainly an XBP and
the orbital period is postulated to be 120~days
(Israel et al. 1998\markcite{Israel1998}).

In K99, RX J0051.3$-$7216 is identified as the ROSAT counterpart
of AX J0051$-$722.
However,
it may be a misidentification
because the separation between
the ROSAT position and ASCA position is $\gtrsim 2^\prime$,
which is significantly larger than the positional uncertainty of ROSAT/ASCA.

\paragraph{No.\,15: 1WGA J0053.8$-$7226}
Corbet et al. (1997a)\markcite{Corbet1997a} discovered
coherent pulsations with a period of 46.63(4)~s
from No.\,15 in observation G.
A ROSAT variable source 1WGA J0053.8$-$7226
and a Be star exist in its error circle
(Buckley et al. 1997\markcite{Buckley1997}),
hence 1WGA J0053.8$-$7226 would be
an XBP with a Be star companion.

\paragraph{No.\,16: XTE J0055$-$724}
An RXTE observation on January 22, 1998 and
a BeppoSAX observation on January 28, 1998 revealed that
XTE J0055$-$724=1SAX J0054.9$-$7226 was exhibiting
coherent pulsations with a 58.969(1)~s period
(Marshall \& Lochner 1998\markcite{Marshall1998}; 
Santangelo et al. 1998\markcite{Santangelo1998}).
ASCA source No.\,16 was detected in the error circle
in observation F.
We performed an epoch folding search and found a weak peak at $\sim 59$~s 
(Figure \ref{fig:efs}),
hence No.\,16 is the counterpart of XTE J0055$-$724.

Israel et al. (1999b)\markcite{Israel1999b} investigated the ROSAT archival data,
and found flux variability of a factor $> 30$.
They also determined the error circle with $10''$ accuracy
within which a Be star was discovered by Stevens et al. (1999)\markcite{Stevens1999}.
XTE J0055$-$724 is thus an XBP with a Be star companion.

\paragraph{No.\,20: AX J0058$-$7203}
Coherent pulsations with a 280.4(4)~s period
from No.\,20 (AX J0058$-$7203) were first detected during this analysis
(Yokogawa \& Koyama 1998b\markcite{Yokogawa1998b}).
Tsujimoto et al. (1999)\markcite{Tsujimoto1999} investigated all 
the Einstein and ROSAT 
observation fields covering the position of AX J0058$-$7203.
They found flux variations with a factor $\gtrsim 10$,
which indicates that
this pulsar is an XBP.

\paragraph{No.\,25: RX J0059.2$-$7138}
The transient source
No.\,25 (RX J0059.2$-$7138) was serendipitously detected
with ROSAT and ASCA on the same day.
Coherent pulsations with a 2.7632(2)~s period were detected from
the ROSAT data
(Hughes 1994\markcite{Hughes1994b}).
We detected the same period in the ASCA data and
hence confirmed the ROSAT result.
Hughes (1994)\markcite{Hughes1994b} proposed a possible optical counterpart
in the error circle, which was later revealed to be a Be star
by Southwell \& Charles (1996)\markcite{Southwell1996}.
These facts indicate that this source is a transient XBP with a
Be star companion.
A broad-band study of this source is presented by
Kohno, Yokogawa, \& Koyama (2000)\markcite{Kohno2000}, 
using the ROSAT and ASCA data simultaneously.

\paragraph{No.\,29: 1SAX J0103.2$-$7209}
Hughes \& Smith (1994\markcite{Hughes1994a}) and
Ye et al. (1995\markcite{Ye1995})
made ROSAT HRI observations of
the shell-like radio SNR 0101$-$724.
No X-ray emission from the radio shell was found,
while a point source
RX J0103.2$-$7209
(=1SAX J0103.2$-$7209)
was detected inside the SNR.
In addition, a Be star has been discovered in the error circle
of RX J0103.2$-$7209.

A BeppoSAX observation on July 27, 1998
discovered coherent pulsations with a 345.2(1)~s period from 1SAX J0103.2$-$7209
(Israel et al. 1998\markcite{Israel1998}),
hence this source could be classified as an XBP with a Be star companion.
Its ASCA counterpart No.\,29 (AX J0103$-$722) has been
observed three times,
in observations B, D, and F.
We detected coherent pulsations with a 348.9(3)~s period
in observation D,
which had the best statistics
(Yokogawa \& Koyama 1998c\markcite{Yokogawa1998c}).
We could not detect coherent pulsations from the data obtained
in the other two observations, probably due to the limited statistics.

The ASCA flux was determined to be
$\sim 1.0 \times 10^{-12}$\uflux\ (2--10~keV)
in each observation.
This value is 3 times smaller than
the BeppoSAX flux
($\sim 3 \times 10^{-12}$\uflux; 2--10~keV).
This variability confirms the binary nature of the source.

\begin{figure}
\psbox[xsize=0.47\textwidth]{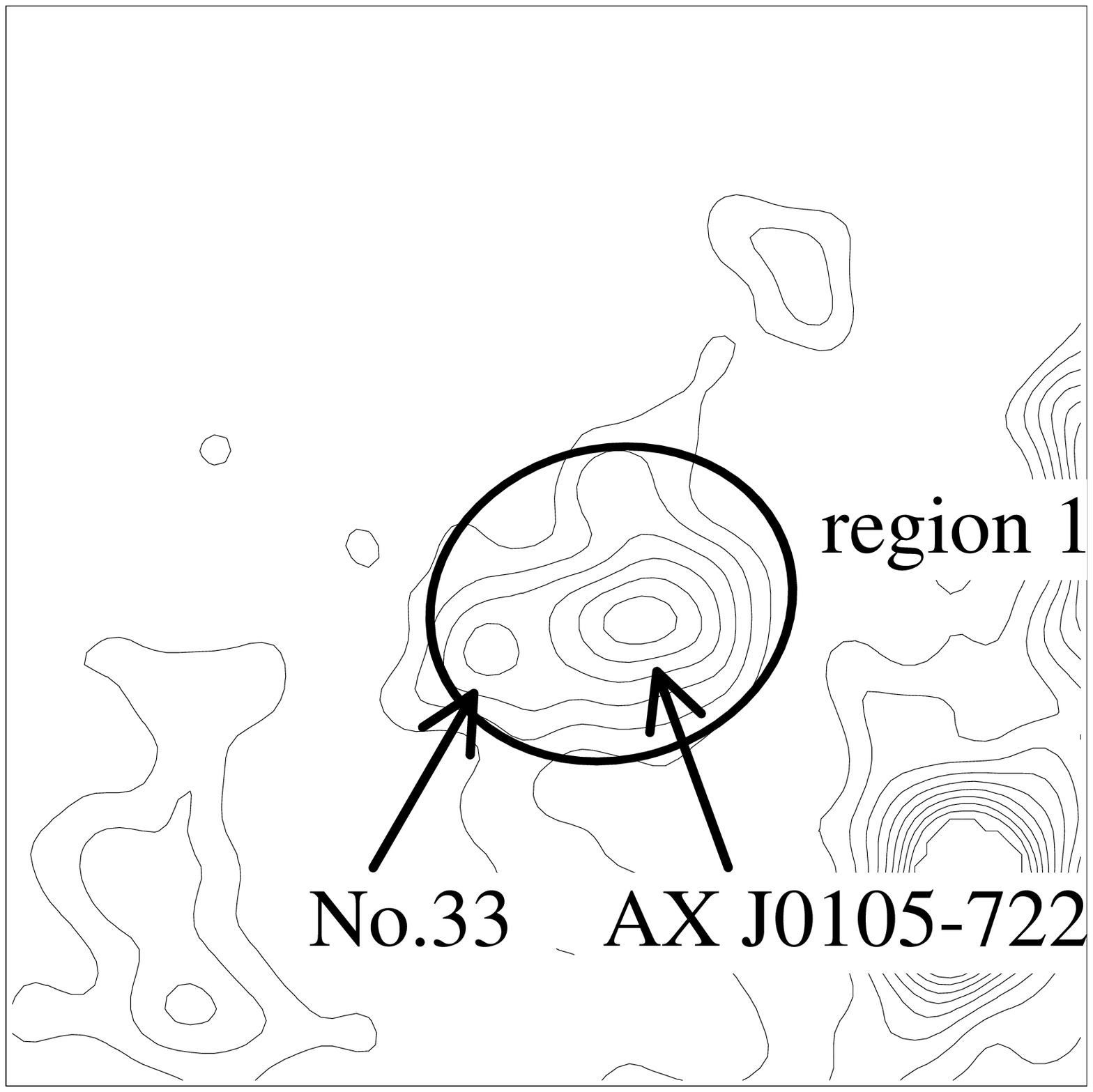}
\psbox[xsize=0.47\textwidth]{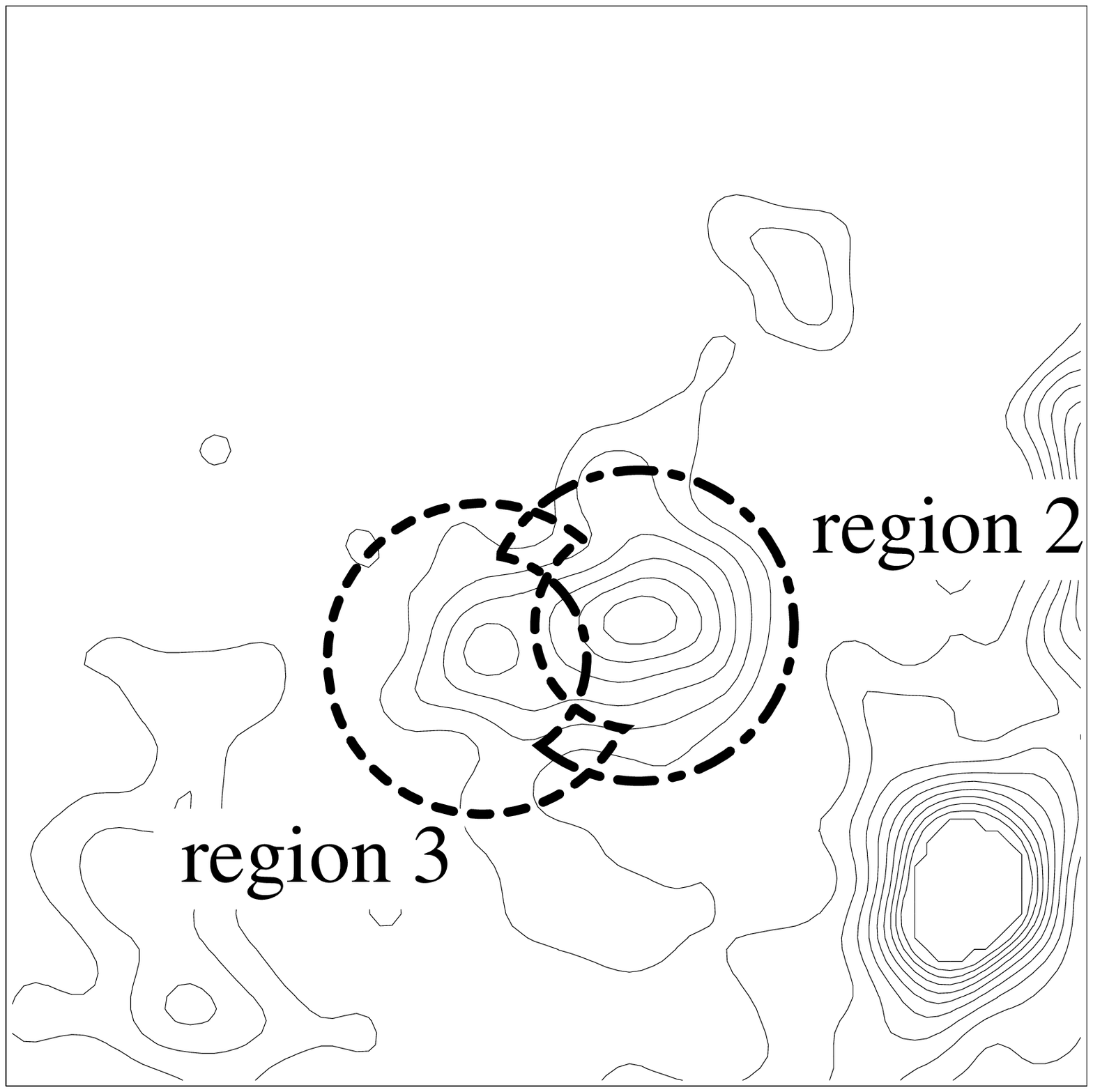}
\caption{X-ray contours obtained by GIS overlaid with
the three regions from which event lists were extracted
to search for coherent pulsations from AX J0105$-$722 (see text).
\label{fig:dems128reg}
}
\end{figure}

\paragraph{No.\,32: AX J0105$-$722 \label{subsec:dems128}}
The separation between the two sources
No.\,32=AX J0105$-$722 and No.\,33 is
only $\sim 3^\prime$.
Using the data in an oval-shaped region
which included both AX J0105$-$722 and No.\,33
(region 1 in Figure \ref{fig:dems128reg}), we
detected coherent pulsations with a 3.34300~s period 
with a marginal significance of $\sim99.5$\%
(Yokogawa \& Koyama 1998d\markcite{Yokogawa1998d}; Figure \ref{fig:psd} (b)).
 Then we separately searched for pulsations from the regions 2 and 3,
shown in Figure \ref{fig:dems128reg}.
We found weak evidence for the 3.34300~s period from region 2
(which includes AX J0105$-$722),
but not from region 3,
hence the pulsations are probably due to AX J0105$-$722.

\placefigure{fig:dems128reg}

The error circle of AX J0105$-$722 includes
a ROSAT source RX J0105.3$-$7210 and
an Einstein source SMC53,
both of which were identified with a radio SNR DEM S128
(Filipovi\'{c} et al. 1998\markcite{Filipovic1998};
WW92\markcite{Wang1992}).
The ASCA flux is
$\sim 2.9 \times 10^{-13}$\uflux\ (0.3--2.4~keV) or
$\sim 3.3 \times 10^{-13}$\uflux\ (0.3--3.5~keV).
These are comparable with fluxes of ROSAT or Einstein
(the discrepancy is within a factor $\sim 2$).

There are several possibilities for the nature of AX J0105$-$722.
Its photon index ($\sim 2.2$), relatively low luminosity
($\sim 10^{35}$\ulumi\ in 0.7--10.0~keV), and the association with an SNR
are common with
SNRs emitting synchrotron X-rays
(e.g. SN1006 --- Koyama et al. 1995\markcite{Koyama1995};
RX J1713.7$-$3946 --- Koyama et al. 1997\markcite{Koyama1997}),
Crab-like SNRs, or
Soft $\gamma$-ray Repeaters
(SGR1806$-$20 --- Kouveliotou et al. 1998\markcite{Kouveliotou1998};
SGR1900$+$14 --- Kouveliotou et al. 1999\markcite{Kouveliotou1999}; 
Murakami et al. 1999\markcite{Murakami1999}; 
Hurley et al. 1999\markcite{Hurley1999}).
Whether AX J0105$-$722 is pulsating or not
and, if it is pulsating, the pulse period and the period derivative
carry essential information
that would enable us to distinguish between
these possibilities.
Further observations with better photon statistics are needed.

\paragraph{No.\,37: XTE J0111.2$-$7317}
XTE J0111.2$-$7317 was serendipitously discovered with RXTE
and CGRO (Compton Gamma-Ray Observatory),
and coherent pulsations with a $\sim 31$~s period were detected
(Chakrabarty et al. 1998a\markcite{Chakrabarty1998a};
Wilson et al. 1998\markcite{Wilson1998}).
Source No.\,37 was detected in the ASCA TOO observation H
(Chakrabarty et al. 1998b\markcite{Chakrabarty1998b}).
Coherent pulsations with a 30.9497(6)~s period were detected,
hence No.\,37 is certainly the counterpart of XTE J0111.2$-$7317.

As described in \S\ref{sec:spec},
this pulsar shows a soft X-ray excess above the simple power-law model.
Yokogawa et al. (1999b)\markcite{Yokogawa1999b} conducted 
a detailed phase-resolved spectroscopic study using both the GIS and SIS,
and concluded that the soft excess is also pulsating with the same pulse
phase and period.

\paragraph{No.\,38: SMC X-1}
SMC X-1 is a well known XBP, having a
B-type supergiant as a companion
(Bildsten et al. 1997\markcite{Bildsten1997}).
During observations A and H,
random flux variations of a factor $\sim 3$ were detected.
In observation C, it was in an eclipse as predicted by the ephemeris 
(Wojdowski et al. 1998\markcite{Wojdowski1998}), 
although we found no clear pulsation of $\sim 0.71$~s (pulse period of SMC X-1)

SMC X-1 has been observed for a long time
and its pulse period is monotonically decreasing
(e.g. Nagase 1989\markcite{Nagase1989};
Wojdowski et al. 1998\markcite{Wojdowski1998};
Kahabka \& Li 1999\markcite{Kahabka1999b}),
including the ASCA observation A.
The period in the observation H is first determined in this study,
and is consistent with the monotonic decrease.

\paragraph{2E 0050.1$-$7247}
Coherent pulsations with a 8.8816(2)~s period
from 2E 0050.1$-$7247=RX J0051.8$-$7231 were discovered by
Israel et al. (1997\markcite{Israel1997}).
Flux variability of a factor 20 and
a Be star in the error circle were detected and
hence this pulsar is an XBP with a Be star companion.
An ASCA observation (G)
covered the position of 2E 0050.1$-$7247.
We did not detect any positive excess above the background level
from the position of this source.
The upper limit of its flux is estimated to be
$\sim 1 \times 10^{-13}$\uflux\ (0.7--10.0~keV),
assuming that the photon index is $\sim 1$.

\paragraph{RX J0052.1$-$7319}
Coherent pulsations with a 15.3~s period from RX J0052.1$-$7319
were discovered in ROSAT and CGRO observations in 1996
(Lamb et al. 1999\markcite{Lamb1999};
Kahabka 1999\markcite{Kahabka1999a}).
Flux variability is also reported, therefore
RX J0052.1$-$7319 could be classified as  an XBP.

The position of No.\,14 coincides with that of RX J0052.1$-$7319.
However, we detected no sign of coherent pulsations
from either an FFT analysis or an epoch folding search.
Therefore, it is not clear whether No.\,14 is really
the counterpart of RX J0052.1$-$7319.

\paragraph{XTE J0054$-$720}
Coherent pulsations were found from XTE J0054$-$720 with RXTE.
Its pulse period and flux in the 2--10~keV band were
169.30~s and $\sim 6.0 \times 10^{-11}$\uflux\
on December 17, 1997, and
168.40~s and $\sim 3.3 \times 10^{-11}$\uflux\
on December 20, 1997,
respectively (Lochner et al. 1998\markcite{Lochner1998a}).
The transient and fading behavior indicates that
XTE J0054$-$720 is an XBP.

A part of its error circle ($\sim 10^\prime$ radius) is
included in the ASCA observations.
Only one source, No.\,17, was detected in the error circle
during ASCA observation F,
although coherent pulsations were not detected.
The flux of No.\,17
($\sim 3.6 \times 10^{-13}$\uflux; 2--10~keV)
is about 100 times smaller than that reported with RXTE.
However, the ROSAT counterpart of No.\,17, RX J0055.4$-$7210, is regarded as
a background source in Filipovi\'{c} et al. (1998)\markcite{Filipovic1998}.

We note that
the ROSAT source RX J0052.9$-$7158 is also in the error circle
of XTE J0054$-$720
(and out of our observation fields).
It was reported to have a large flux variability
and a Be star companion (Cowley et al. 1997\markcite{Cowley1997}),
and hence may be the counterpart of XTE J0054$-$720.

\paragraph{RX J0117.6$-$7330}
RX J0117.6$-$7330 was serendipitously discovered
in a ROSAT PSPC observation on September 30--October 2, 1992
(Clark, Remillard, \& Woo 1996\markcite{Clark1996}).
The luminosity was $2.3\times10^{37}$\ulumi\
between 0.2--2.5~keV at that time
(Clark, Remillard, \& Woo 1997\markcite{Clark1997}),
and was found to diminish by a factor of over 100 within one year.
Macomb et al. (1999)\markcite{Macomb1999}
discovered coherent pulsations with a $\sim 22.07$~s period
from the same data,
with the aid of archival data obtained by BASTE onboard CGRO
in the same epoch.
Strong Balmer emission lines and IR excess were detected
from the companion star in the error circle 
(Coe et al. 1998)\markcite{Coe1998}, 
indicating that RX J0117.6$-$7330 is an XBP with a Be star companion.

Although the position of RX J0117.6$-$7330 was covered in
ASCA observations A and C, no X-ray emission was detected.
It was difficult to estimate the upper limits of the flux
because of the contamination from SMC~X-1
located only $\sim 5'$ away from RX J0117.6$-$7330.

\placefigure{fig:pulspec}
\placefigure{fig:pullc}

\section{Discussion \label{sec:dis}}

\subsection{Classification by Hardness Ratio \label{subsec:classification}}

 We have classified many of the SMC sources into
two classes (XBPs and thermal SNRs), with the aid of
their temporal/spectral properties and other information.
Since the number of thermal SNRs is only four, 
more samples of thermal SNRs are essential 
to make any statistical discussion with XBPs and thermal SNRs.
X-ray sources in the LMC is suitable for the inclusion 
for the reason described in the next subsection, 
thus we included 
XBPs and thermal SNRs in the LMC from all the available ASCA data
(40 observations as of 1998).
LMC X-4,
EXO 053109$-$6609.2,
1SAX J0544.1$-$710, and
A0538$-$67
have been detected with ASCA,
and are regarded as XBPs due to
their long-term flux variability
and optical counterparts 
(Bildsten et al. 1997\markcite{Bildsten1997};
Burderi et al. 1998\markcite{Burderi1998};
Cusumano et al. 1998\markcite{Cusumano1998};
Haberl, Dennerl, \& Pietsch 1995\markcite{Haberl1995};
Corbet et al. 1997b\markcite{Corbet1997b}).
Although we could not detect coherent pulsations from A0538$-$67,
this source is very likely to be an XBP
with a Be star companion (Corbet et al. 1997b\markcite{Corbet1997b},
and references therein).
Ten SNRs in the LMC,
N103B, 0509$-$67.5, 0519$-$69.0,
N23, N49, N63A, DEM71, N132D, 0453$-$68.5, and N49B,
have been found to exhibit emission lines from ionized atoms
(Hughes et al. 1995\markcite{Hughes1995};
Hayashi 1997\markcite{Hayashi1997};
Hughes, Hayashi, \& Koyama 1998\markcite{Hughes1998}),
hence are classified as thermal SNRs.
In addition, ASCA has observed the SNRs
N44, N86, 0548$-$70.4, 0534$-$69.9, and 0543$-$68.9.
We analyzed their spectra to search for any emission lines
as described in \S\ref{sec:spec},
and consequently N44 and 0548$-$70.4 were classified as thermal SNRs.
We also included two other established classes:
Crab-like SNRs\footnote{In this section,
``Crab-like SNR'' indicates a SNR associated with
a rotation-powered X-ray pulsar.}
(0540$-$69 and N157B)
and BHs (LMC X-1 and LMC X-3).

We derived HRs and fluxes of the four XBPs, 12 thermal SNRs,
two Crab-like SNRs and two BHs in the LMC.
To estimate their fluxes,
we fitted the GIS spectra with simple models;
an absorbed power-law model for
XBPs, Crab-like SNRs, and BHs,
while
an absorbed thin-thermal CIE plasma model for thermal SNRs.
In the thermal plasma model,
the global abundance was fixed to that
of the interstellar matter in the LMC,
0.3~solar (Russell \& Dopita 1992\markcite{Russell1992};
Hayashi 1997\markcite{Hayashi1997};
Hughes et al. 1998\markcite{Hughes1998})
for simplicity.
The line profiles of some thermal SNRs
could not be reproduced by the simple thermal plasma model.
However, the continuum shapes were well described and
hence the derived fluxes are unlikely to have any large systematic error.
To estimate the error, we fitted the GIS spectra of
two thermal SNRs in the SMC,
0102$-$723 and 0103$-$726,
with the same simple model
where the SMC abundance was assumed for all elements.
The best-fit models
well traced the continuum shapes,
showing the same discrepancy in the line profiles.
Derived fluxes between 0.7--10.0~keV were
$1.3 \times 10^{-11}$\uflux\ (0102$-$723) and
$1.1 \times 10^{-12}$\uflux\ (0103$-$726),
respectively,
while detailed analysis in \S\ref{subsec:snrs}
yielded fluxes of
$1.4 \times 10^{-11}$\uflux\ (0102$-$723) and
$1.1 \times 10^{-12}$\uflux\ (0103$-$726),
respectively.
Therefore the systematic flux error caused by the simple model
is estimated to be $\lesssim 10$\%.

\begin{figure}
\psbox[xsize=0.97\textwidth]{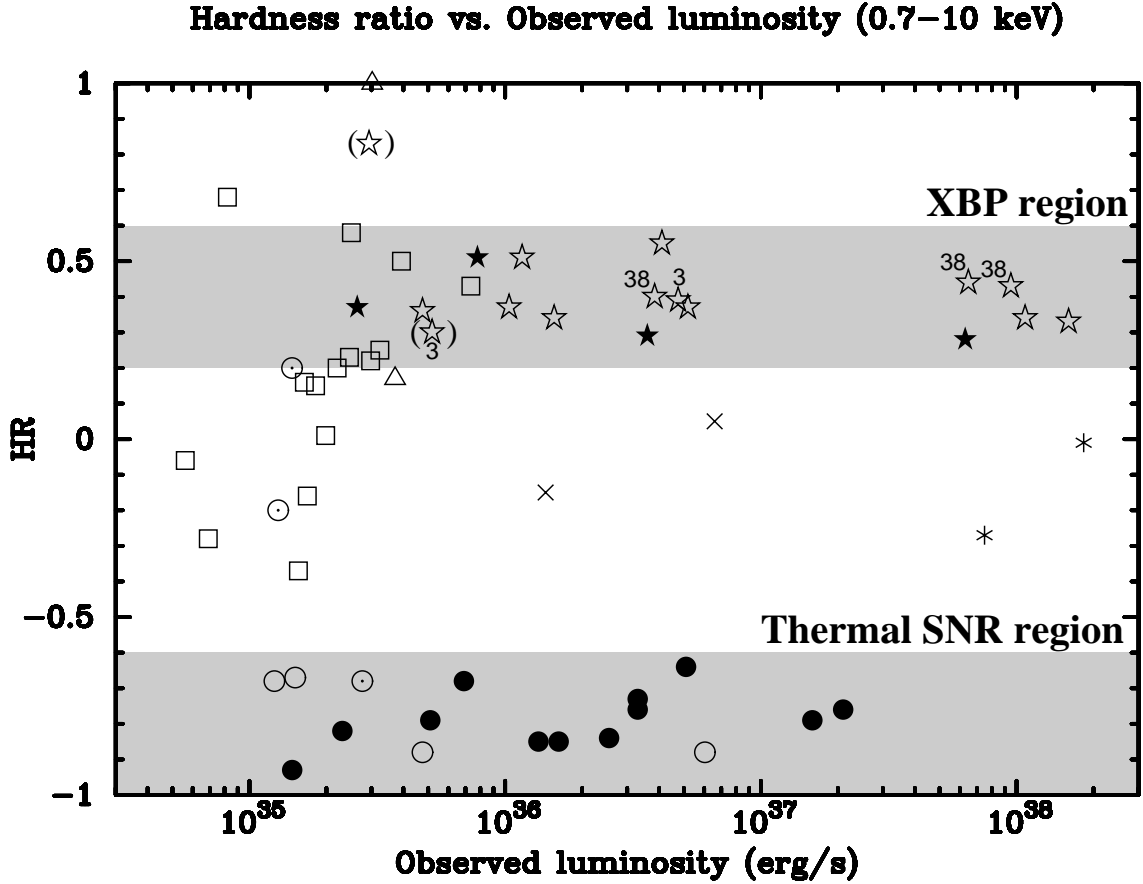}
\caption{Plot of HR as a function of the observed luminosity
in 0.7--10.0~keV.
Open/filled symbols represent the SMC/LMC sources.
Symbols represent
XBPs (stars), thermal SNRs (circles),
Crab-like SNRs (crosses), BHs (asterisks),
other radio SNRs (dotted circles),
non-pulsating HMXBs (triangles) and
unclassified sources (squares), respectively.
Typical error of HR of
the latter three classes is
$\sim 0.2$.
 No.\,39 (Galactic star HD 8191) is omitted here. 
The variable sources detected multiple times 
are marked with the source numbers 
(No.\,3=AX J0049$-$729 and No.\,38=SMC X-1).
Stars in parentheses are 
positionally coincident with known XBPs, 
from which we detected no coherent pulsations
(AX J0049$-$729 and RX J0052.1$-$7319). 
\label{fig:hr-lobs_smclmc_class_unid}
}
\end{figure}

In Figure \ref{fig:hr-lobs_smclmc_class_unid},
HRs of the sources in these classes are plotted against the observed luminosity $L_{\rm obs}$,
defined as $L_{\rm obs} = F_{\rm x} \times 4\pi d^2$,
where $d$ is the distance to the SMC (60~kpc) or LMC (50~kpc).
The LMC samples (represented by filled symbols)
largely enhanced the number of thermal SNRs,
and we found a clear split between XBPs and thermal SNRs.
Therefore we can safely say that
the HRs of all the XBPs in the SMC and LMC
fall in a narrow region of $0.2\leq{\rm HR}\leq0.6$
(``XBP region''),
while those of all the thermal SNRs
fall in a region of $-1.0\leq{\rm HR}\leq-0.6$
(``thermal SNR region'').

\placefigure{fig:hr-lobs_smclmc_class_unid}

\subsection{Validity of Our Classification Method}

 Source classification with HR has also been
carried out with ROSAT data (K99).
The common characteristic of the method of K99 and ours is
that the HR is defined independently of any model:
only the number of photons detected by ROSAT/ASCA detectors is used.
Hence their/our HR is
an ``apparent hardness of the spectrum.''
Such a simple method is not valid for X-ray sources in our Galaxy,
because soft X-rays are absorbed by interstellar matter
which gives a dependence of the HR on source distance.

 To check this, we simulated HRs for simple spectra
either of thin-thermal plasma with a temperature of 0.5~keV,
a typical value for a thermal SNR, or power-law with a
photon index of 1.0, a typical value for an XBP.
 When the absorption column density is $3\times 10^{21}$~H~cm$^{-2}$,
which roughly corresponds to a distance of 1~kpc, HRs are
$-$0.89 and 0.36, for the thermal and power-law model, respectively.
 For a larger absorption column density of $8\times 10^{22}$~H~cm$^{-2}$,
which is equivalent to the Galactic center region,
the HR of the thermal model or the power-law model
increased to be 0.5 or 0.96, respectively.
 We therefore conclude that in our Galaxy
the HR values can have a large scatter even for the same class,
depending on the source distances.
On the other hand, according to HI observations made by Luks (1994), 
we can safely assume that the interstellar absorption is rather uniform
towards the SMC and LMC with a scale down to $15'$ (HPBW). 
In fact, HI column density derived by  Luks (1994) was 
$\sim 1 \times 10^{20}$--$7 \times 10^{21}$~H~cm$^{-2}$. 
Within this $N_{\rm H}$ range, the HR values vary very small:
from $-$0.94 to $-$0.81 (thin-thermal spectra) or from 0.25 to 0.48 
(power-law spectra). Thus we conclude that possible uncertainty of HR 
due to the $N_{\rm H}$ spatial variation is smaller than the width of 
the XBP region and thermal SNR region, 
hence the present classification by HR value should be reliable 
for X-ray sources in the Magellanic Clouds. 

Our proposed method is much more simple compared with that of K99;
the classification criteria in K99 require
two hardness ratios and source extent, but
our method needs only one hardness ratio.

\subsection{Candidate of XBPs and Thermal SNRs \label{subsec:cand}}

 In the HR-luminosity plane (Figure \ref{fig:hr-lobs_smclmc_class_unid}), 
no thermal SNR is found in the XBP region,
nor vice versa. In addition, other classes (Crab-like SNRs and BHs)
are not found in these two regions.
Hence we can safely select candidates for XBPs or for thermal SNRs
using HRs.
Among the unclassified sources, shown by squares
in Figure \ref{fig:hr-lobs_smclmc_class_unid},
we find eight XBP candidates and one thermal SNR candidate.
Note that the number of candidates is tentative,
because the HRs of low-flux sources have large errors ($\sim 0.2$).
 We detected no pulsations from the eight XBP candidates,
possibly due to their limited statistics (low fluxes).
 The thermal SNR candidate is the radio SNR 0049$-$736,
which was not classified as thermal in \S\ref{sec:spec}
because of the limited photon statistics.
We note that the other unclassified SNRs
DEM S128 (No.\,32) and 0056$-$725 (No.\,21)
are out of the thermal SNR region.
This fact confirms the possible non-thermal nature
described in \S\ref{subsec:snrs}.
 In addition, 0056$-$725 is an XBP candidate and
hence X-rays from this source may be attributed to
an (unresolved) XBP in the radio SNR 0056$-$725,
as 1SAX J0103.2$-$7209 (No.\,29).

\subsection{Source Populations in the SMC \label{sec:pop}}

 As described in \S\ref{sec:pul} and \S\ref{subsec:cand},
we have 14 XBPs and eight candidates in the SMC.
 With a typical exposure time of 40 ksec for
the present SMC and LMC observation, we
can obtain reasonable photon statistics
for the sources brighter than
$\sim 1\times10^{-12}$\uflux\ (0.7--10.0~keV),
which corresponds to
$\sim 4\times10^{35}$\ulumi.
 Therefore our observations cover the luminosity ranges
of typical HMXBs and LMXBs.

 Since most XBPs are HMXBs
(e.g. Bildsten et al. 1997\markcite{Bildsten1997}),
we assume all of the XBPs and the candidates to be HMXBs
for simplicity.
 In addition, at least 11 X-ray sources have been identified
with high mass stellar optical counterparts, even though
coherent pulsations have not been detected (summarized in Table \ref{tab:hmxbcat}).
Adding these, we can count
33 HMXBs in the SMC.
On the other hand, about 70 HMXBs are listed in our Galaxy
(van Paradijs 1995\markcite{Paradijs1995};
Bildsten et al. 1997\markcite{Bildsten1997};
Nagase 1999\markcite{Nagase1999}).

\placetable{tab:hmxbcat}

No LMXB has ever been discovered in the SMC
including in this work,
while in our Galaxy, about 100 LMXBs have been found so far
(van Paradijs 1995\markcite{Paradijs1995}).

 The radio surveys of SNRs are more complete than those of X-rays at this moment,
and 14 and 220 radio SNRs are cataloged in the SMC and our Galaxy,
respectively
(Mathewson et al. 1983\markcite{Mathewson1983};
Mathewson et al. 1984\markcite{Mathewson1984};
Mathewson et al. 1985\markcite{Mathewson1985b};
Filipovi\'{c} et al. 1998\markcite{Filipovic1998};
Green 1998\markcite{Green1998}).
Among them, no Crab-like SNR is known in the SMC, while seven are
known in our Galaxy (e.g. Nagase 1999\markcite{Nagase1999};
Pavlov, Zavlin, \& Tr\"{u}mper 1999\markcite{Pavlov1999}).
The thermal SNR candidate defined in \S\ref{subsec:cand}
is the radio SNR 0049$-$736, hence the number of SNRs in the SMC remains unchanged.
In our Galaxy, new X-ray SNRs have been found with ROSAT and ASCA, thus
over 100 X-ray SNRs are expected to be cataloged in the near future
(e.g. Aschenbach 1996\markcite{Aschenbach1996}).

 The estimated numbers of these classes in the SMC and our Galaxy
are summarized in Table \ref{tab:pop}.
Since the mass ratio of our Galaxy to the SMC is roughly 100,
the source numbers in the SMC should be multiplied by 100
for a simple comparison with our Galaxy.
Since hard X-rays from HMXBs brighter than $\sim 5\times10^{35}$\ulumi\
can penetrate the interstellar gas through the entire Galaxy, we may neglect
any selection effect caused by Galactic absorption.
Thus we conclude that number of HMXBs normalized by the galaxy
mass is much higher in the SMC than in our Galaxy.
Accordingly, we suspect that the SMC has been
more active than our Galaxy in massive star formation.

\placetable{tab:pop}

The number ratio of HMXBs to LMXBs exhibits
a striking difference as has already
been pointed out by Schmidtke et al. (1999)\markcite{Schmidtke1999}.
Our results make this difference even larger,
especially when we include the HMXB candidates; 33/0 in the SMC and 70/100 in our Galaxy.
LMXBs are generally considered to comprise an older population
than HMXBs, because of their properties such
as spatial distribution outside
active star forming regions and weak magnetic fields
(e.g. Lewin, van Paradijs, \& Taam 1995\markcite{Lewin1995}).
Accordingly, the high ratio in the SMC implies
rather recent star formation activity about several $10^{\rm 6-7}$~yr ago.

Both Crab-like SNRs and HMXBs originate from massive stars:
the former from a single star, while the latter from a binary system.
The number ratios of these classes in our Galaxy (7/70)
and that in the SMC (0/33) are largely different from each other
in spite of the similar origin.
This may imply that the formation efficiency of massive binary stars
relative to that of single massive stars was much higher in the SMC than in our Galaxy.

\section{Conclusion \label{sec:conc}}

 We have performed systematic analyses of X-ray sources in the SMC
with all the available data obtained with ASCA.

Among the 16 X-ray pulsars in the SMC,
five were newly discovered during our analyses.
We also confirmed the pulsations of another seven pulsars.
We regarded 14 pulsars out of the 16 as XBPs
because of their flux variability
and, for some sources, the existence of an optical counterpart.

Eight radio SNRs were detected.
The spectral analyses revealed that
at least four SNRs exhibit the emission lines of ionized atoms,
hence they are expected to be thermal SNRs.
Another two SNRs are possibly emitting non-thermal X-rays.

We established a simple and reliable method of
source classification using the hardness ratio (HR).
XBPs fall in a range of $0.2\leq {\rm HR} \leq0.6$,
while thermal SNRs fall in a range of $-1\leq {\rm HR}\leq -0.6$.
 With this method,
eight XBP candidates and one thermal SNR candidate
are found.
 Normalized by the galaxy mass, HMXBs are found
to be extremely populous in the SMC, compared with those in our Galaxy.
 An even more striking contrast between the two galaxies is the number ratio
of HMXBs to LMXB; it is 33/0 in the SMC but is 70/100 in our Galaxy.
The lack of Crab-like SNRs in the SMC may indicate that the
formation of single massive stars in the SMC was less active than
that of binary systems.

\acknowledgments
The authors express their thanks to all the members of the ASCA team. 
This research has made use of data obtained through 
the High Energy Astrophysics Science Archive Research Center Online Service,
provided by the NASA/Goddard Space Flight Center.
J.Y. and M.N. are financially supported by the Japan Society for
the Promotion of Science.

\onecolumn

\begin{table}
\caption{SMC Fields Observed with ASCA. \label{tab:obs}}
\end{table}

\begin{table}
\caption{First ASCA Catalog of X-ray Sources in the SMC.
\label{tab:smccat}}
\end{table}

\begin{table}
\caption{X-ray Pulsars in the SMC.\label{tab:pcat}}
\end{table}

\begin{table}
\caption{Best-Fit Parameters of 0103$-$726 for the CIE or NEI Model.
\label{tab:spec0103}}
\end{table}

\begin{table}
\caption{Non-Pulsating HMXBs in the SMC.
\label{tab:hmxbcat}}
\end{table}

\begin{table}
\caption{X-ray Source Populations in the SMC and Our Galaxy.
\label{tab:pop}}
\end{table}

\newpage

\begin{figure}
\psbox[xsize=0.47\textwidth]{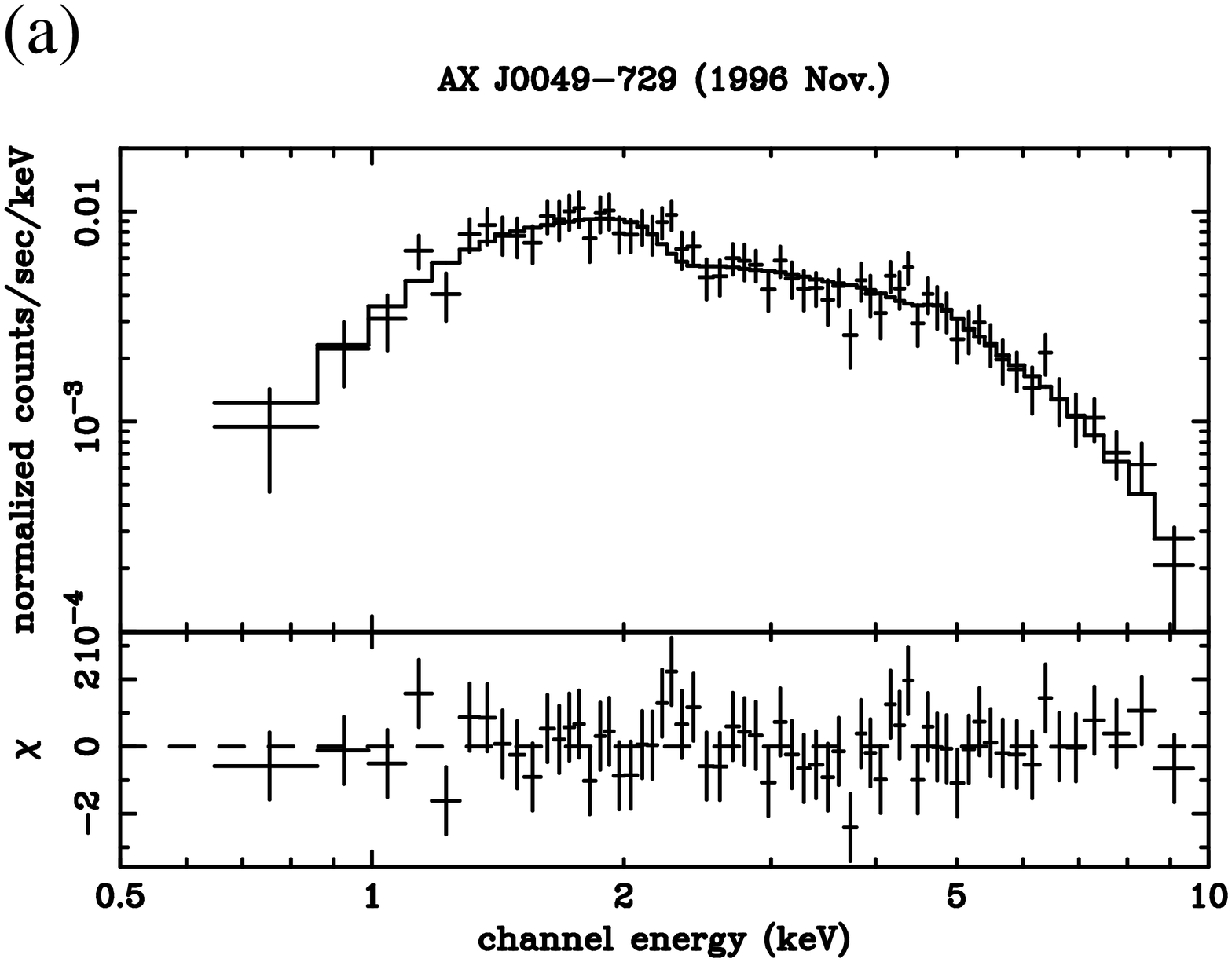}
\psbox[xsize=0.47\textwidth]{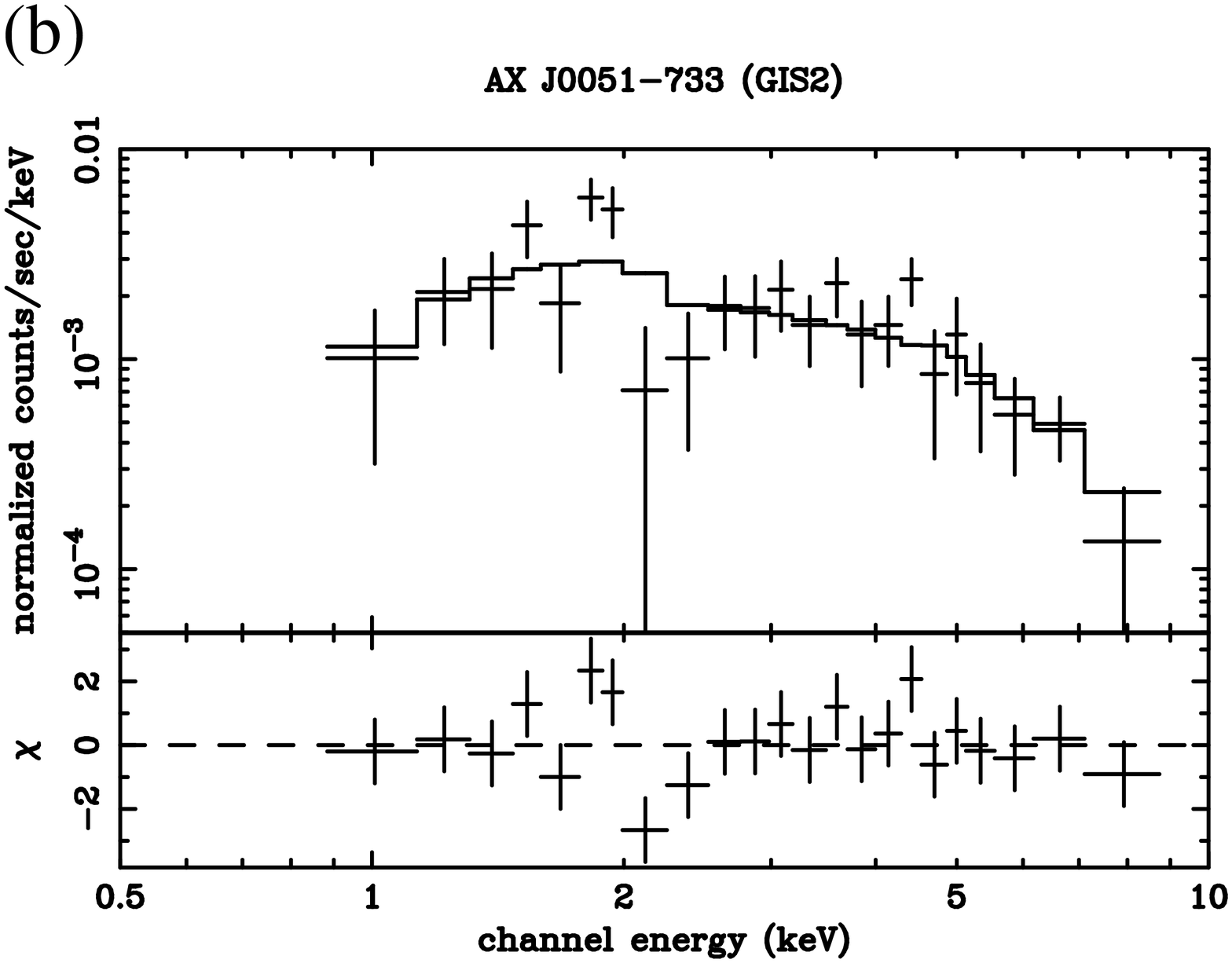}

\psbox[xsize=0.47\textwidth]{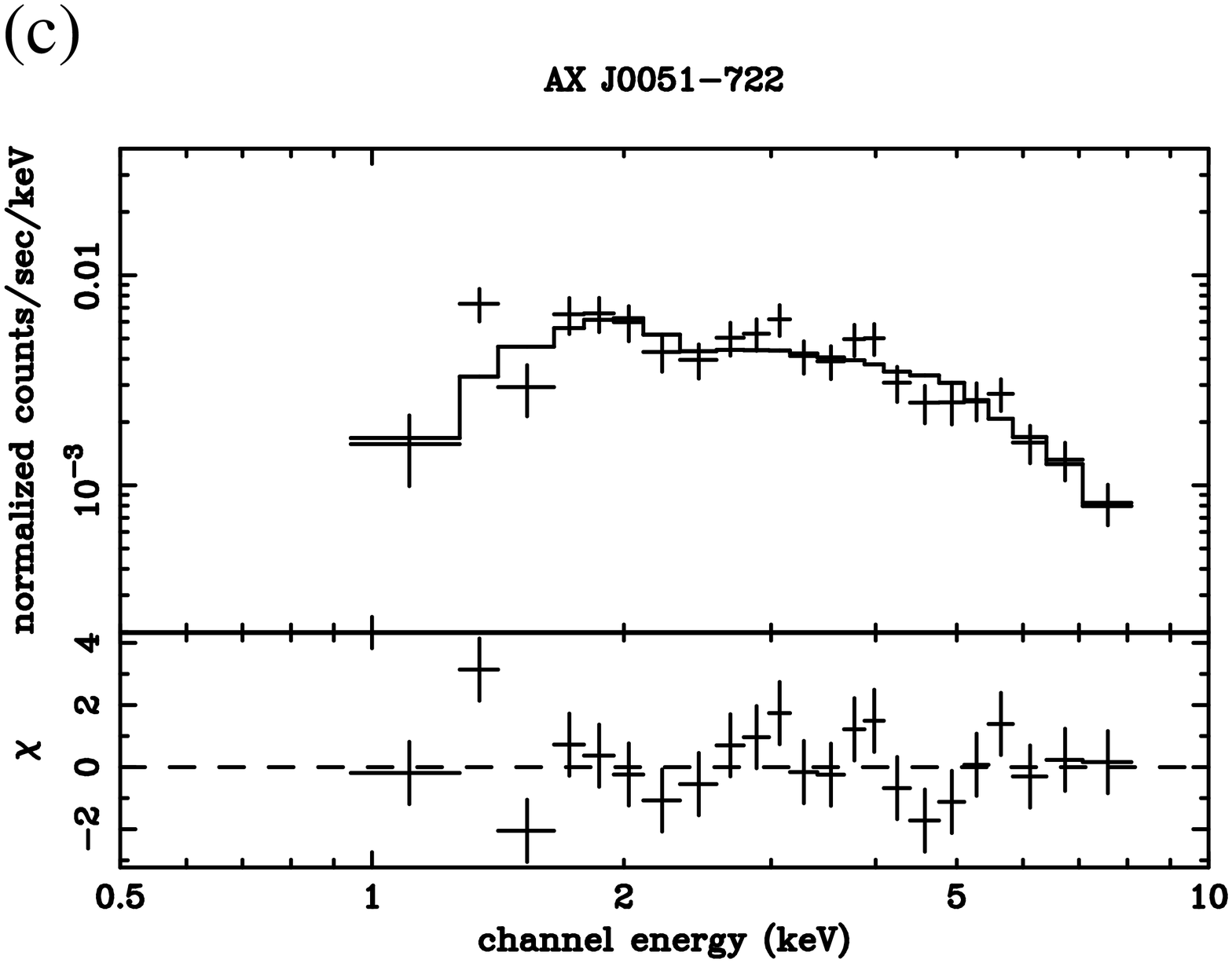}
\psbox[xsize=0.47\textwidth]{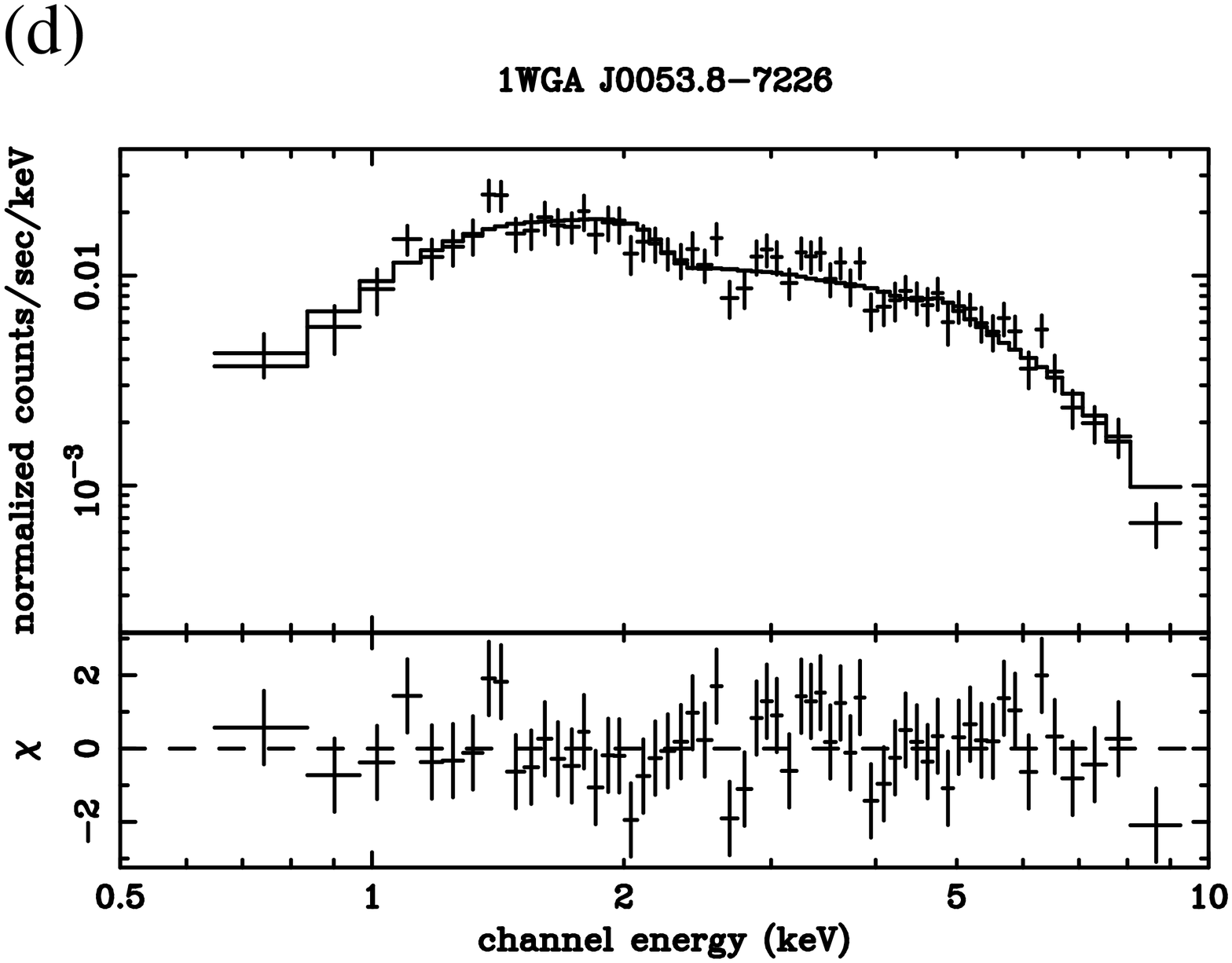}

\psbox[xsize=0.47\textwidth]{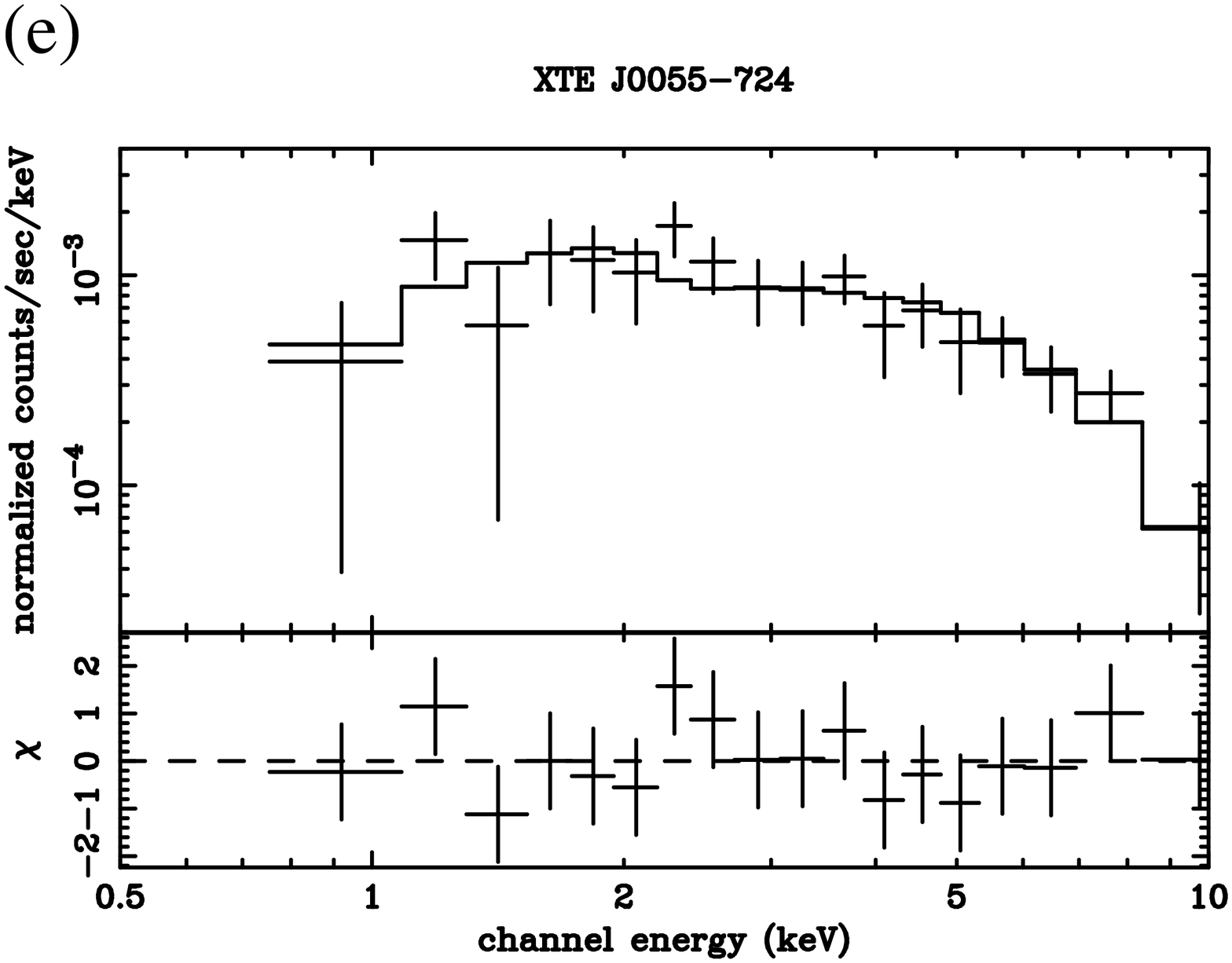}
\psbox[xsize=0.47\textwidth]{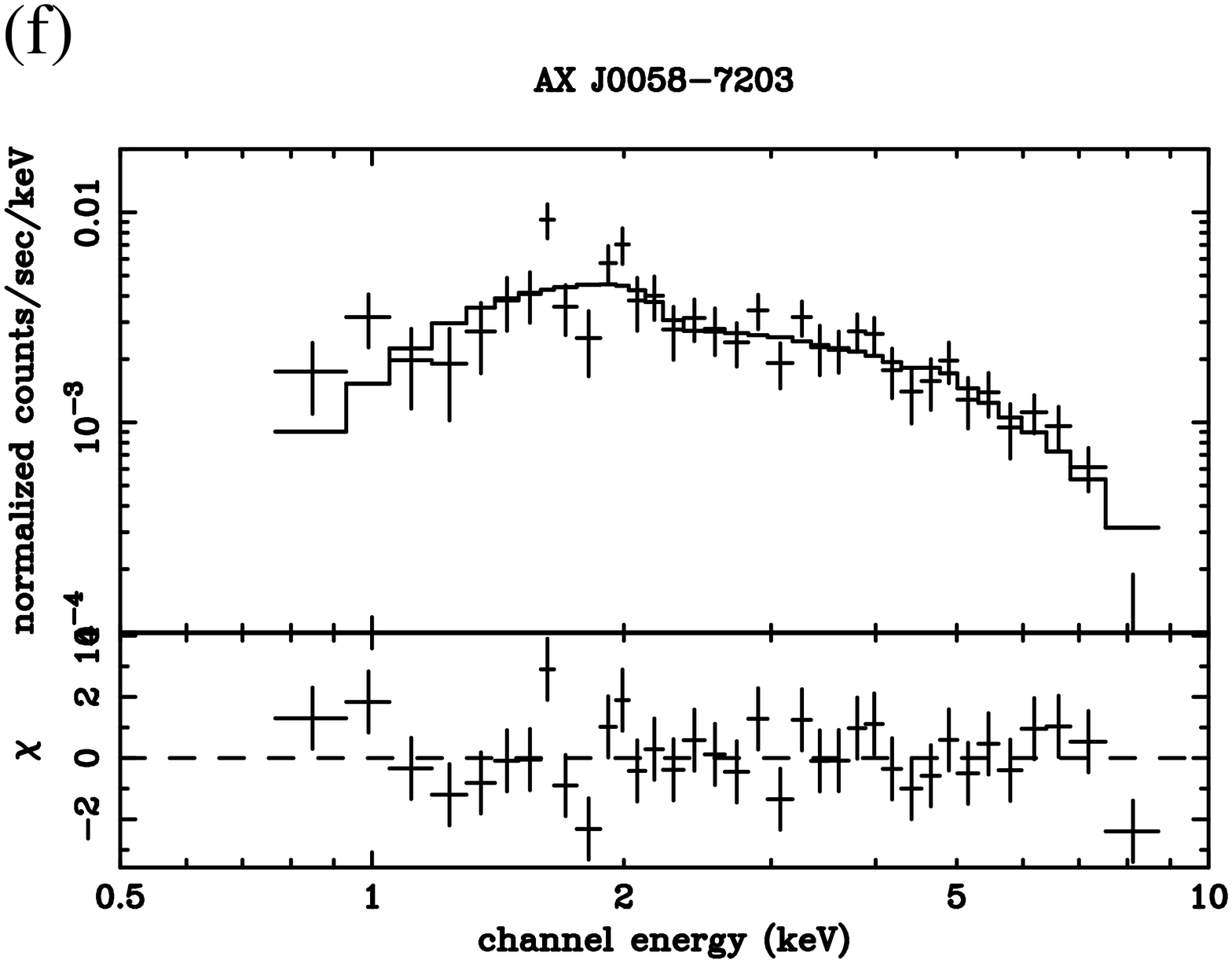}

\psbox[xsize=0.47\textwidth]{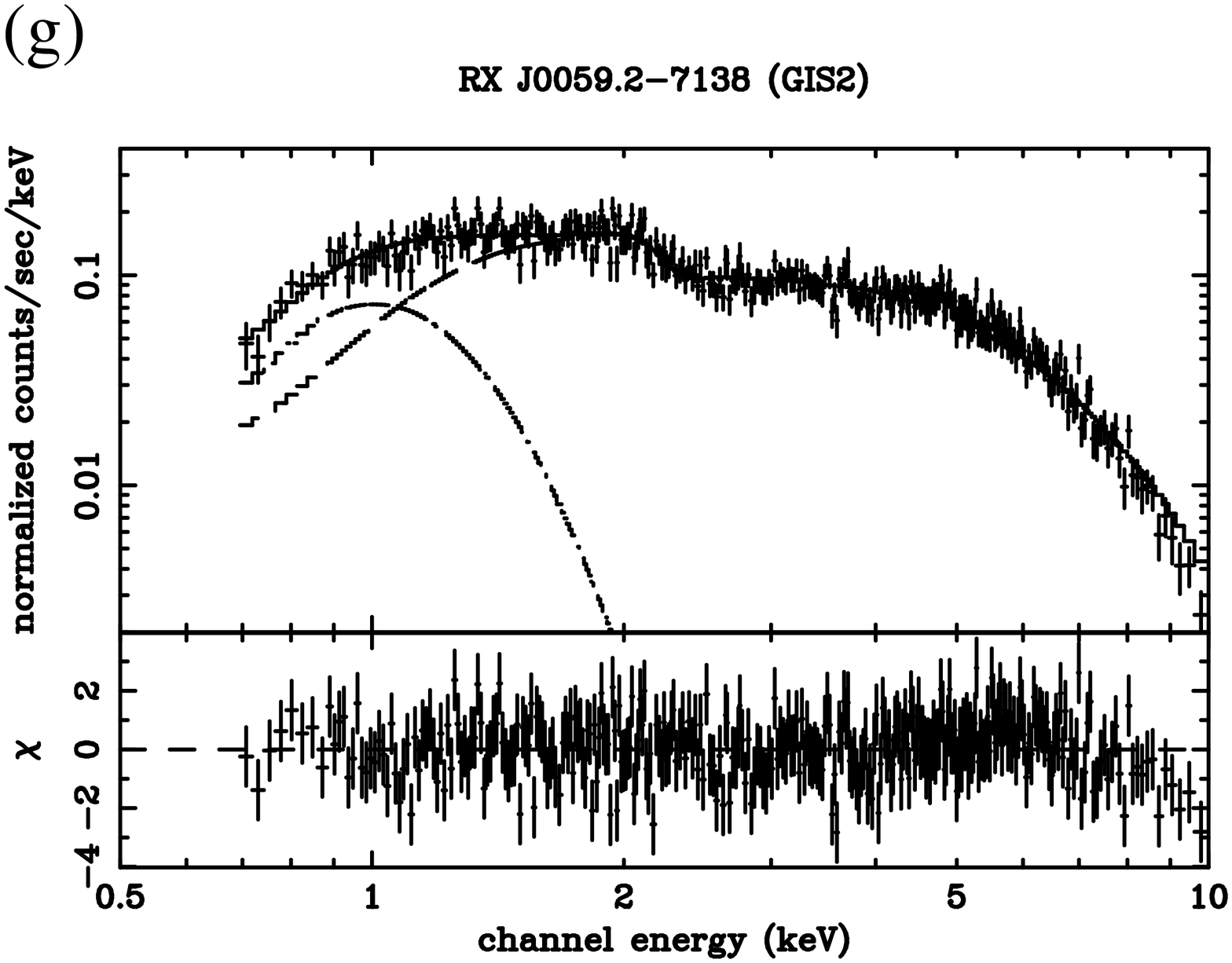}
\psbox[xsize=0.47\textwidth]{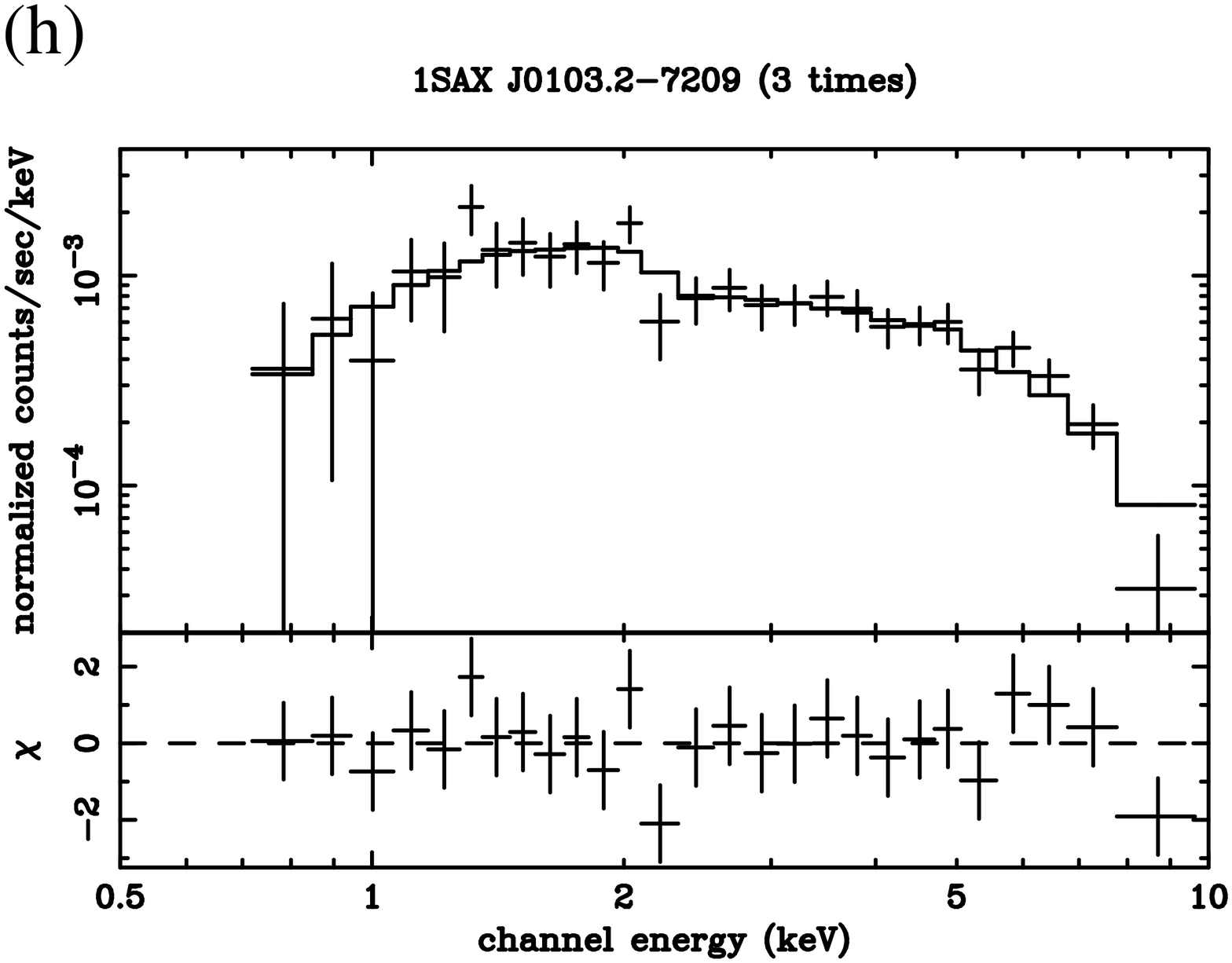}
\end{figure}

\begin{figure}
\psbox[xsize=0.47\textwidth]{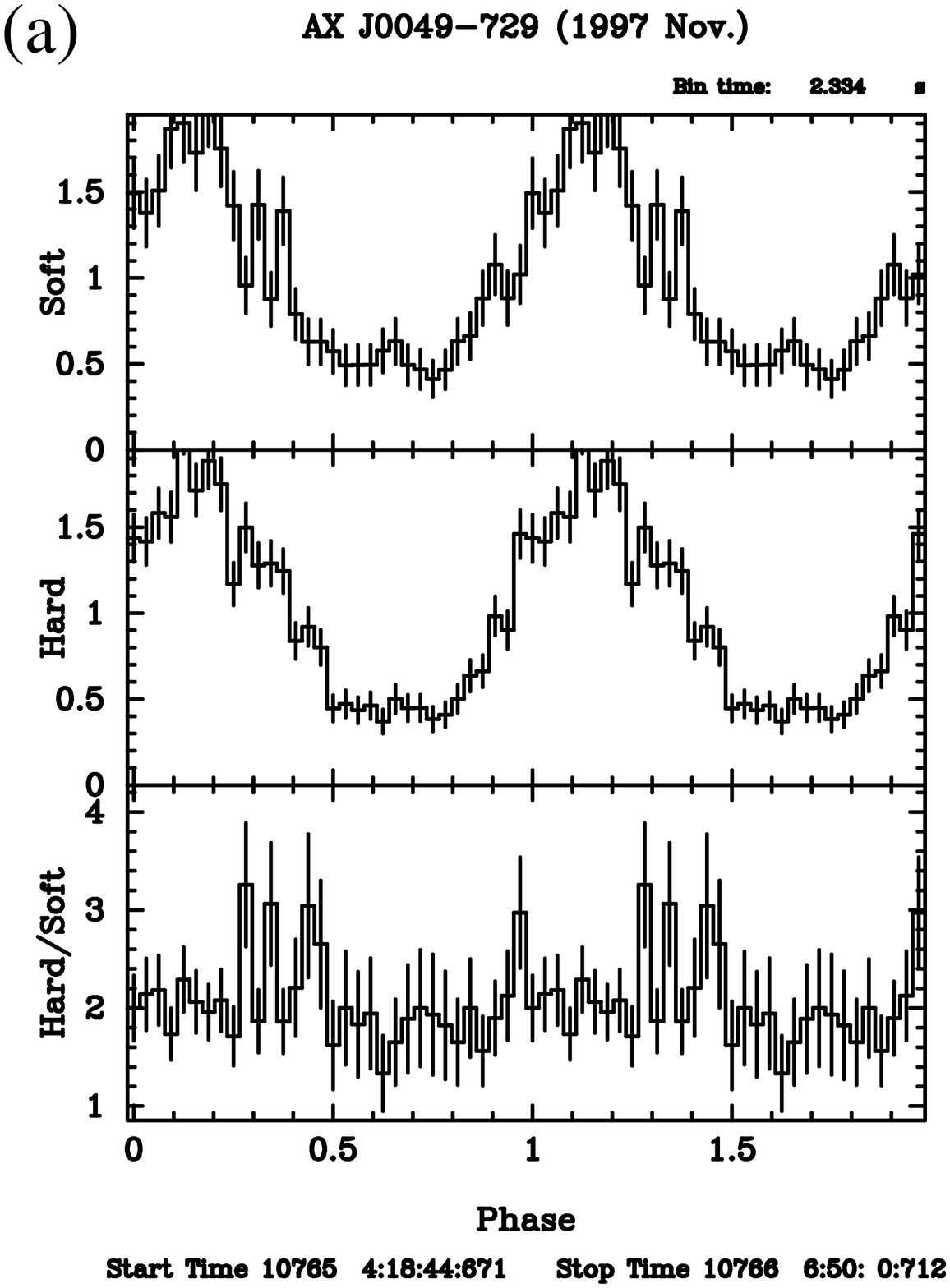}
\psbox[xsize=0.47\textwidth]{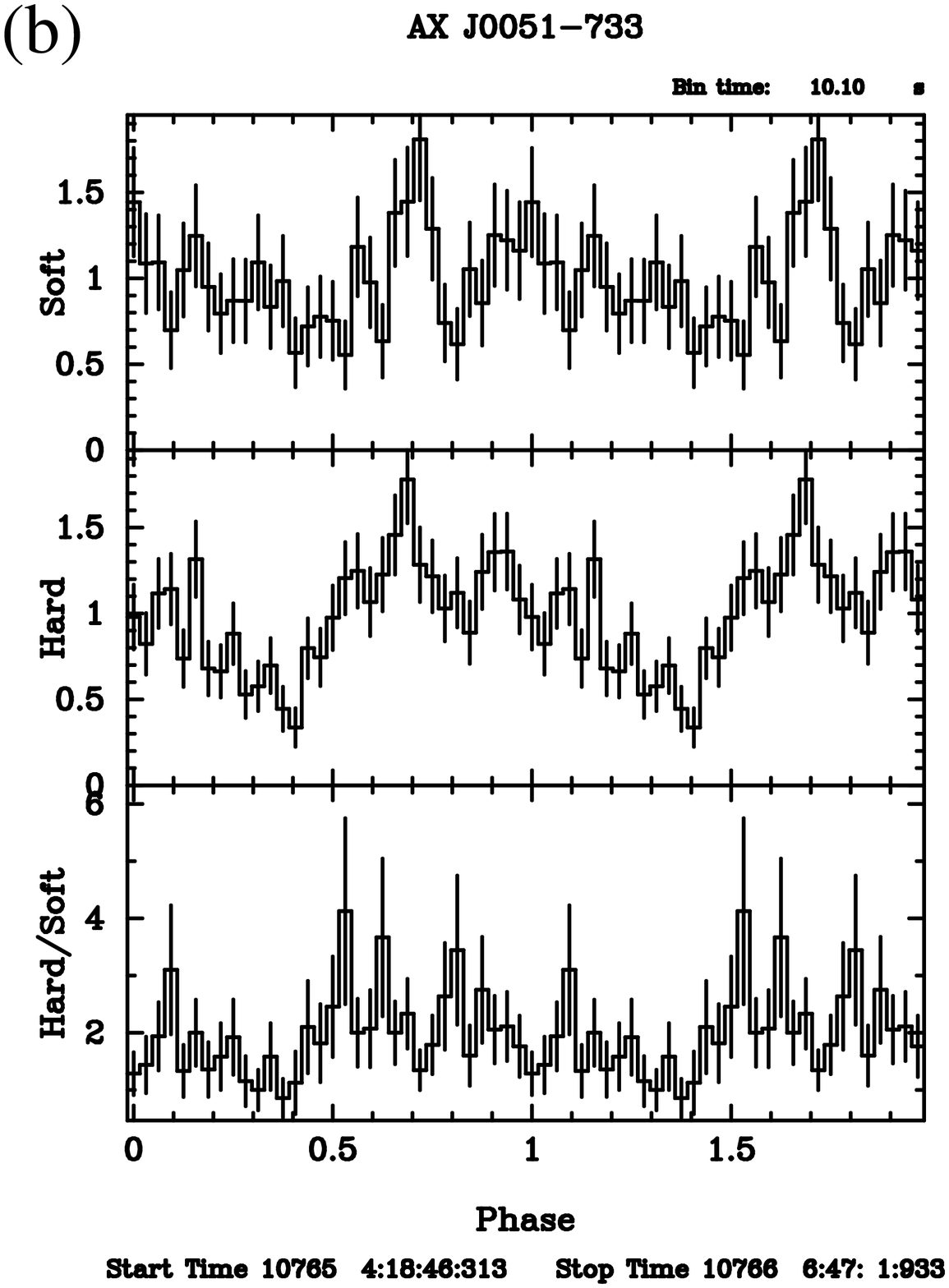}

\psbox[xsize=0.47\textwidth]{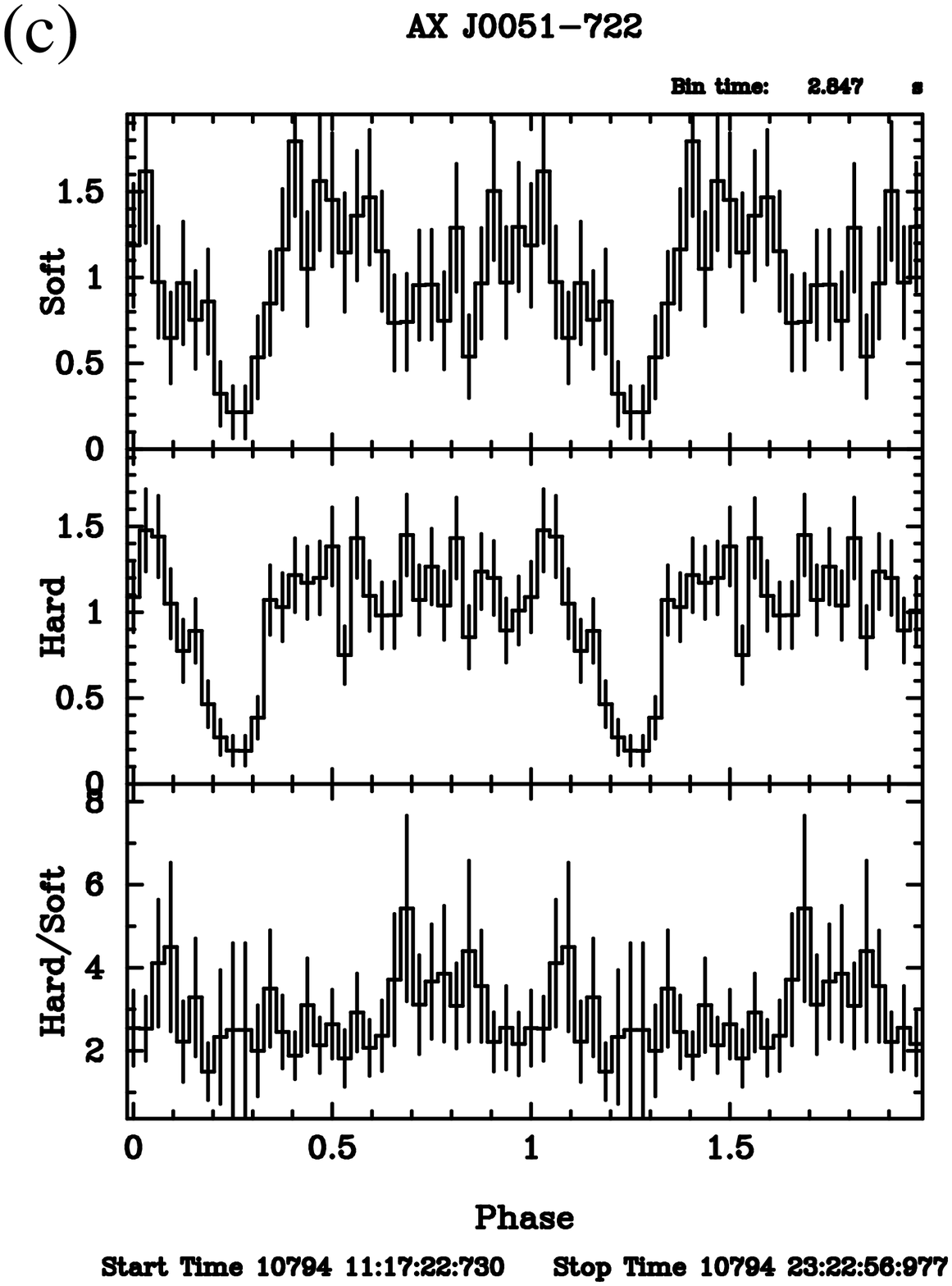}
\psbox[xsize=0.47\textwidth]{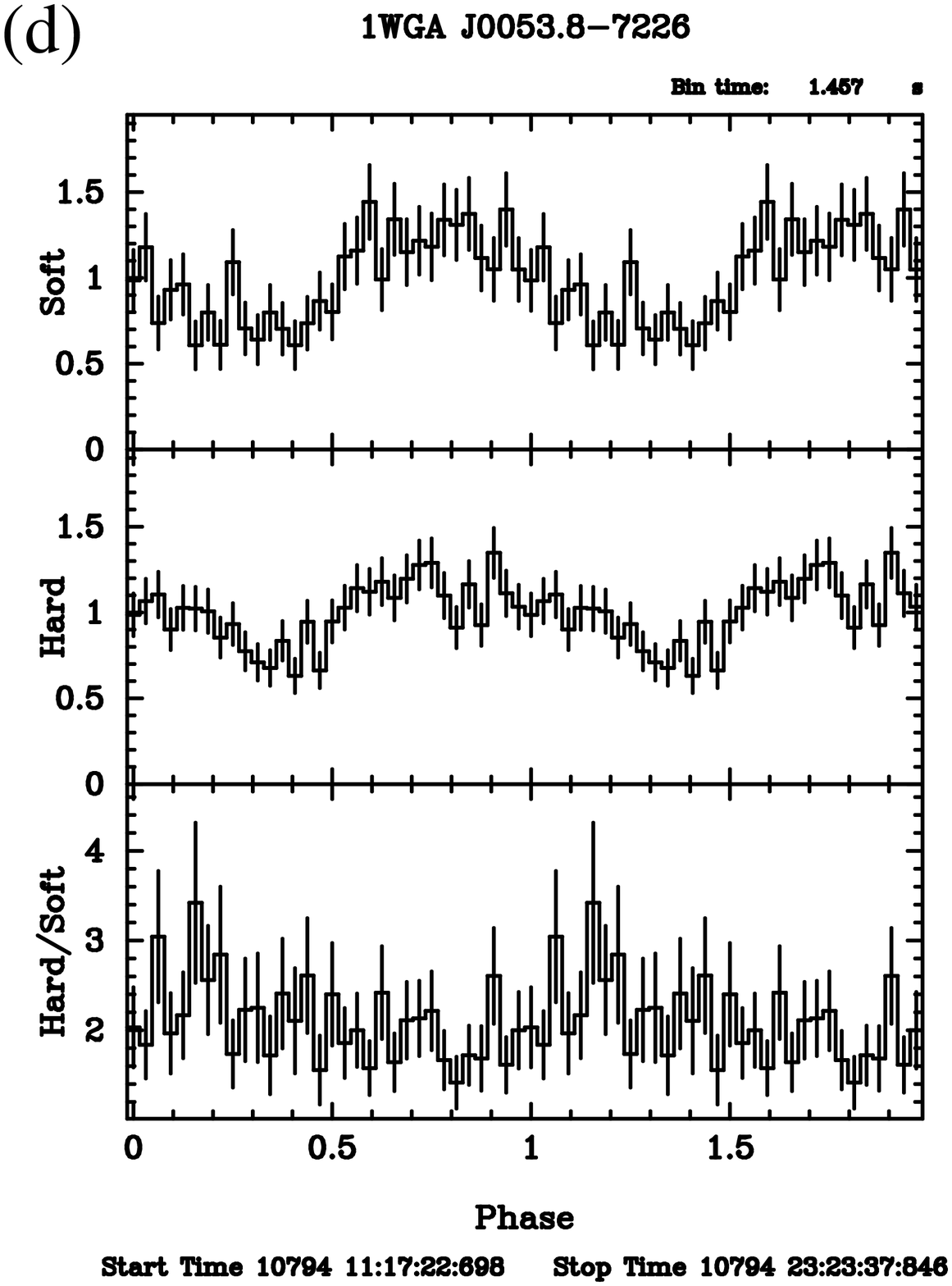}
\end{figure}

\begin{figure}
\psbox[xsize=0.47\textwidth]{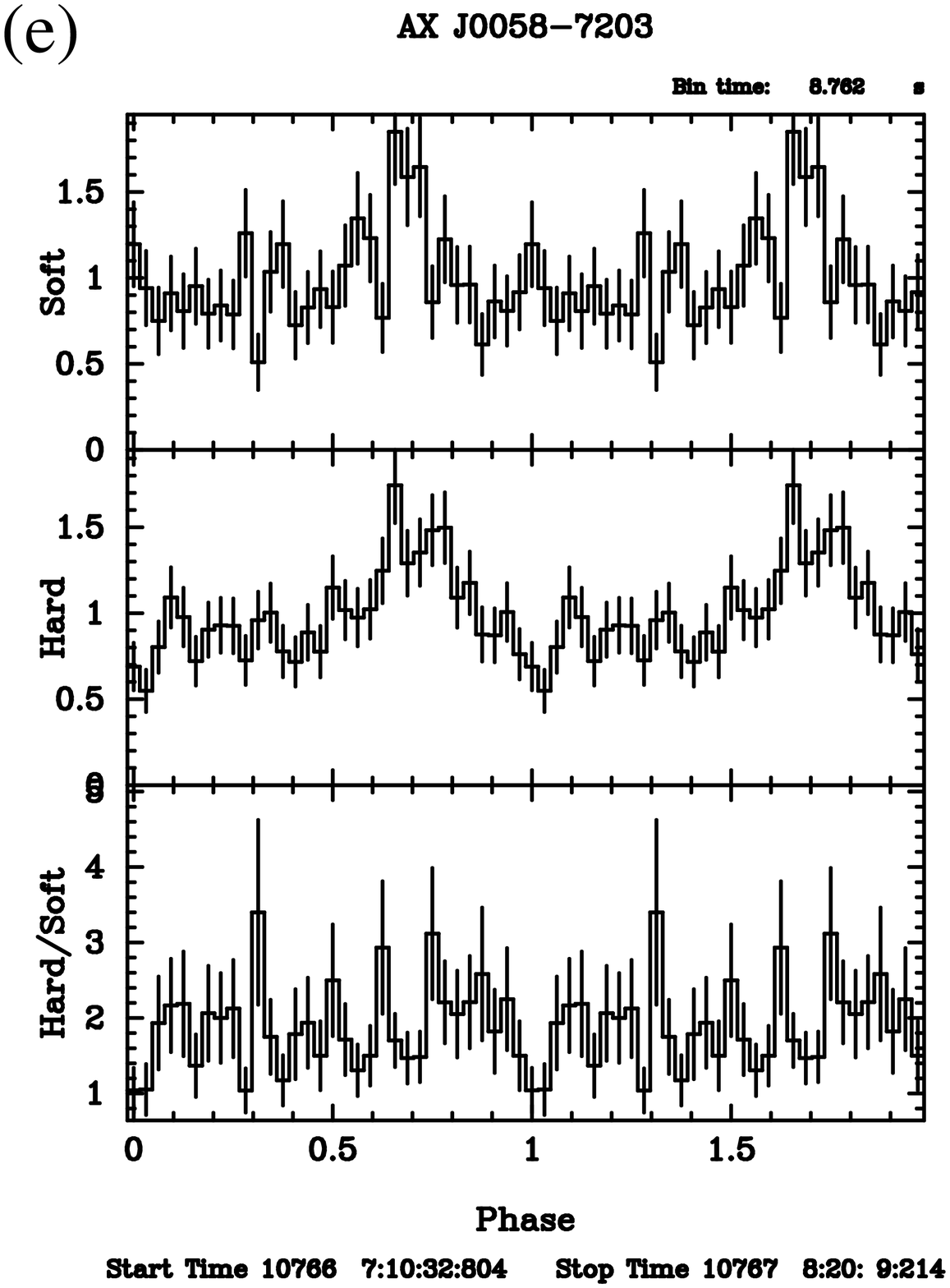}
\psbox[xsize=0.47\textwidth]{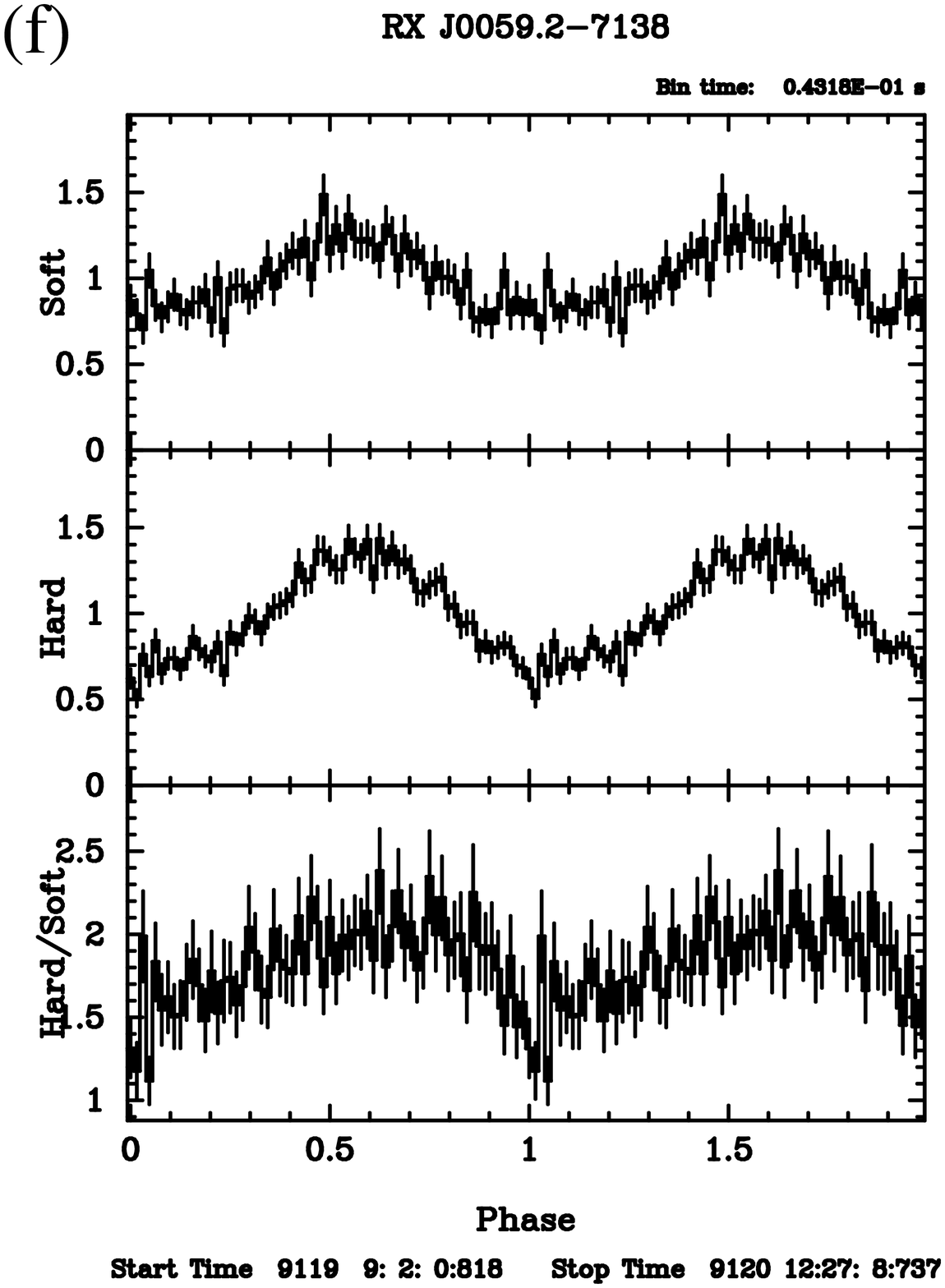}

\psbox[xsize=0.47\textwidth]{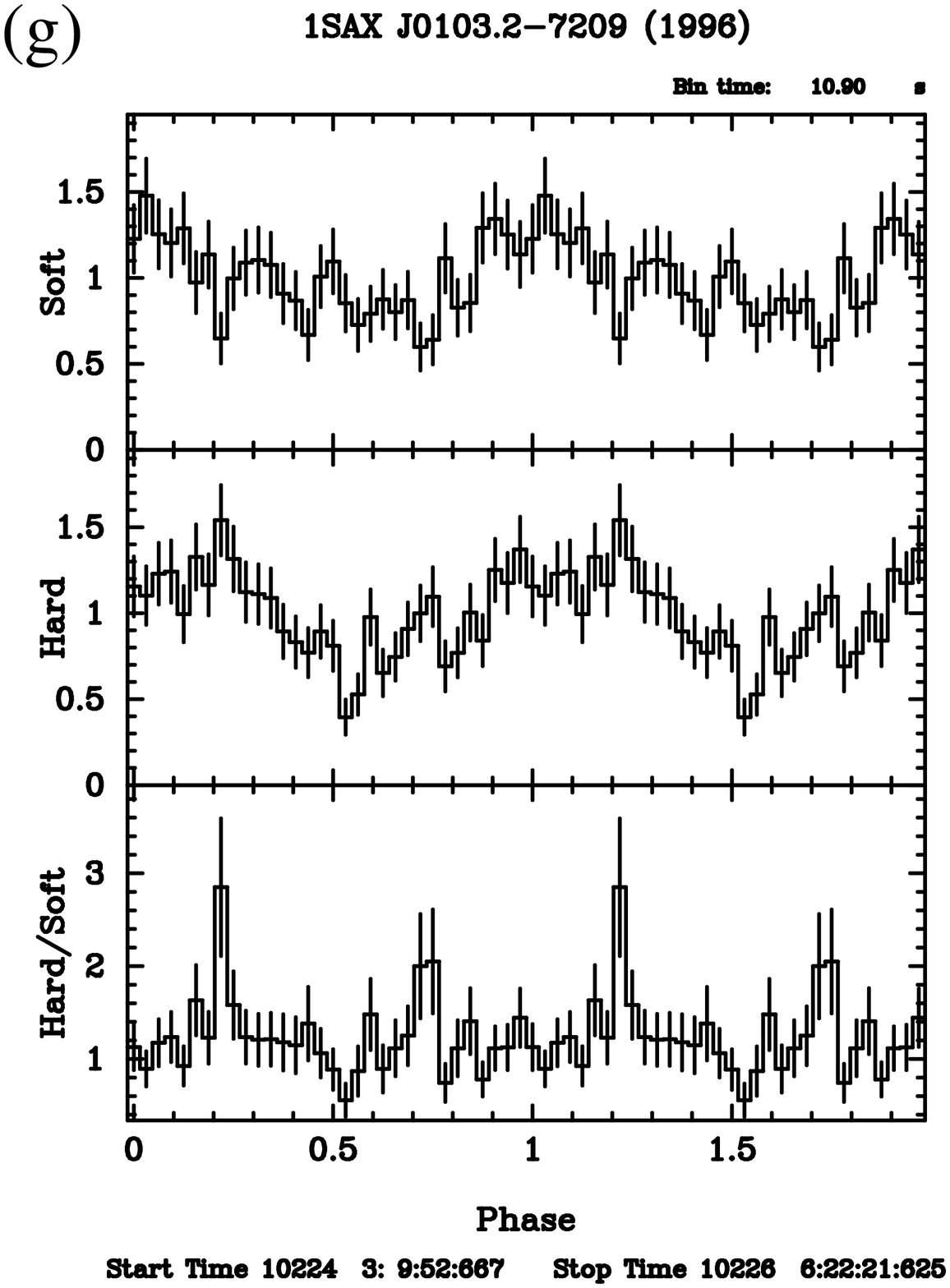}
\psbox[xsize=0.47\textwidth]{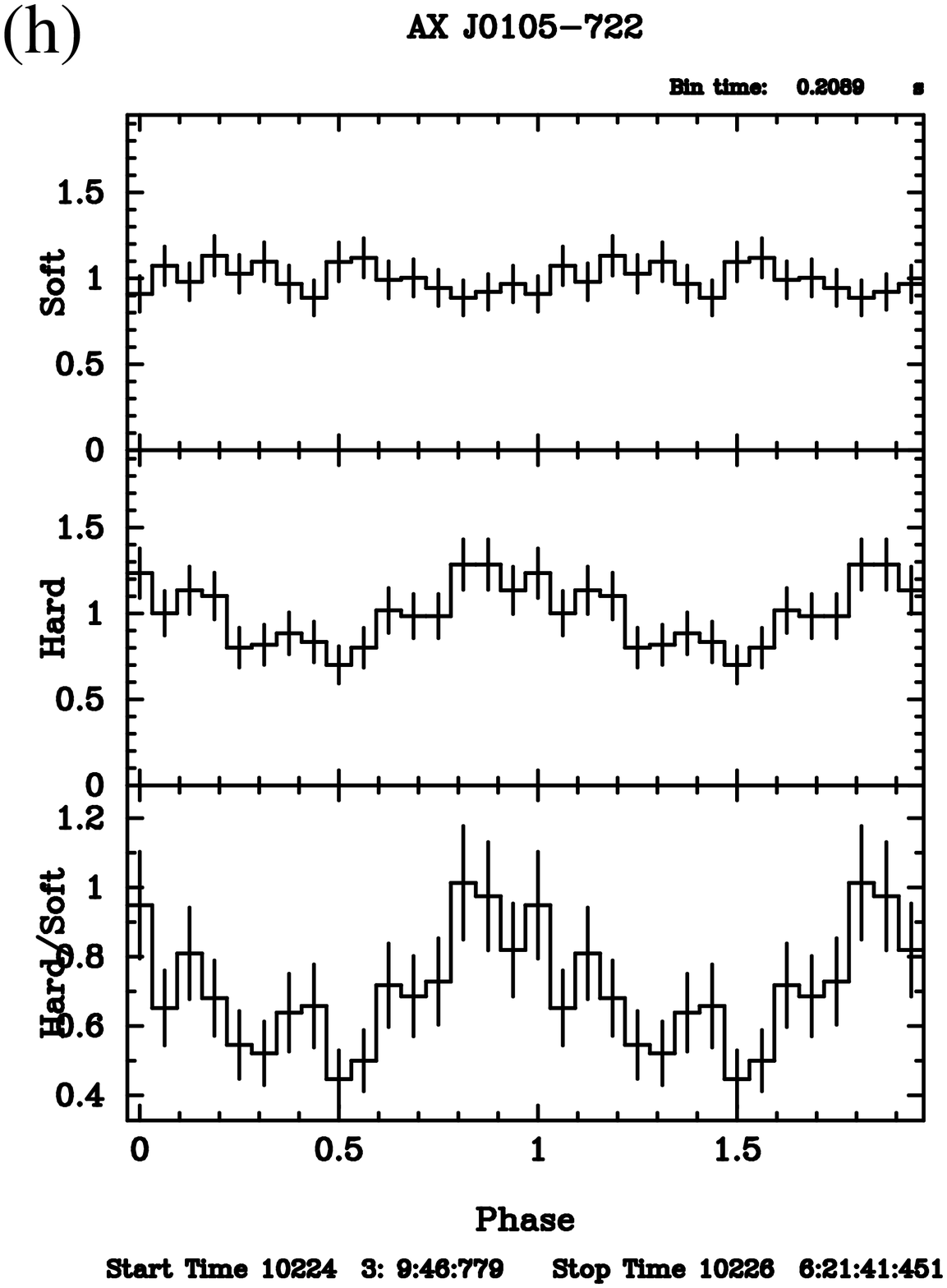}
\end{figure}

\end{document}